\documentclass[lettersize,journal]{IEEEtran}
\usepackage{amsmath,amsfonts}
\usepackage[ruled,linesnumbered]{algorithm2e}
\usepackage{array}
\usepackage{textcomp}
\usepackage{stfloats}
\usepackage{url}
\usepackage{verbatim}
\usepackage{graphicx}
\usepackage{cite}
\usepackage[labelformat=simple]{subcaption}

\usepackage{color}
\usepackage[colorlinks,linkcolor = black,anchorcolor = black,citecolor = black]{hyperref}
\usepackage{xspace}

\begin{document}
\newcommand{\name}{DeepStream\xspace}

\title{\name: Prototyping Deep Joint Source-Channel Coding for Real-Time Multimedia Transmissions}
\author{Kaiyi Chi,
Yinghui He,
Qianqian Yang,~\IEEEmembership{Member,~IEEE},
Zhiping Jiang,
Yuanchao Shu,~\IEEEmembership{Senior Member,~IEEE},
Zhiqin Wang,
Jun Luo,~\IEEEmembership{Fellow,~IEEE}, and
Jiming Chen,~\IEEEmembership{Fellow,~IEEE}

\thanks{Kaiyi Chi, Qianqian Yang, Yuanchao Shu, and Jiming Chen are with the State Key Laboratory of Industrial Control Technology, Zhejiang University, Hangzhou 310027, China (e-mail: \{chikaiyi17, qianqianyang20, ycshu, cjm\}@zju.edu.cn).}
\thanks{Yinghui He and Jun Luo are with the College of Computing and Data Science, Nanyang Technological University, Singapore 639798 (email:\{yinghui.he, junluo\}@ntu.edu.sg).}
\thanks{Zhiping Jiang is with the School of Computer Science and Technology, Xidian University, Xi’an 710126, China (e-mail: zpj@xidian.edu.cn).}
\thanks{Zhiqin Wang is with the School of Information and Communication Engineering, Beijing University of Posts and Telecommunications, Beijing 100876, China, and also with China Academy of Information and Communications Technology, Beijing 100191, China (e-mail: zhiqin.wang@caict.ac.cn).}
}



\maketitle

\begin{abstract}
    Deep learning-based joint source-channel coding (DeepJSCC) 
    has emerged as a promising technique in 6G for enhancing the efficiency and reliability of data transmission across diverse modalities, particularly in low signal-to-noise ratio (SNR) environments.
    This advantage is realized by leveraging powerful neural networks to learn an optimal end-to-end mapping from the source data directly to the transmit symbol sequence, eliminating the need for separate source coding, channel coding, and modulation.
    Although numerous efforts have been made towards efficient DeepJSCC,
    they have largely stayed at 
    numerical simulations that can be far from practice, leaving the real-world viability of DeepJSCC largely unverified.
    To this end, we \textit{prototype} \name upon orthogonal frequency division multiplexing (OFDM) technology to offer efficient and robust 
    DeepJSCC
    for multimedia transmission.
    In conforming to OFDM, we develop both a feature-to-symbol mapping method and a cross-subcarrier precoding method
    to improve the subcarrier independence
    and reduce peak-to-average power ratio. 
    To reduce system complexity
    and enable flexibility 
    in accommodating varying quality of service requirements,
    we further propose a progressive coding strategy that adjusts the compression ratio based on latency with minimal performance loss.
    We implement \name for real-time image transmission and video streaming using software-defined radio. Extensive evaluations verify that \name outperforms both the standard scheme and the direct deployment scheme. Particularly, at an SNR of 10~\!dB, \name achieves a PSNR of 35~\!dB for image transmission and an MS-SSIM of 20~\!dB for video streaming, whereas the standard scheme fails to recover meaningful information.
\end{abstract}

\begin{IEEEkeywords}
Joint source-channel coding, semantic communications, prototyping, multimedia transmissions.
\end{IEEEkeywords}

\maketitle
\vspace{-1.0em}

\section{Introduction}
The upcoming sixth-generation (6G) wireless networks are envisioned to support data-intensive applications with extreme reliability and ubiquitous connectivity~\cite{duan20236g}. In specific use cases such as augmented reality (AR) and unmanned aerial vehicles (UAVs), the volume of data to be transmitted is substantial, posing significant challenges for conventional communication systems~\cite{chen20235g, wang2023road}. For example, remote UAV control systems that rely on high-definition video streaming require highly efficient and robust transmission techniques~\cite{3gpp22125}. As a result, the demand for effective transmission of multimedia data across various modalities has increased exponentially. Recent advances in deep learning (DL) have enabled the development of DL-based codecs and communication systems that leverage the powerful feature extraction and representation capabilities of neural networks~\cite{jain20sec, jiang21mobicom, lu23aaai}. These systems have demonstrated strong potential in significantly improving transmission efficiency, particularly under challenging wireless conditions~\cite{ballé2018variational, shao2021learning}.

\begin{figure}[b]
    \vspace{-1.5em}
    \centerline{\includegraphics[width=0.99\linewidth, trim= 20 5 15 0, clip]{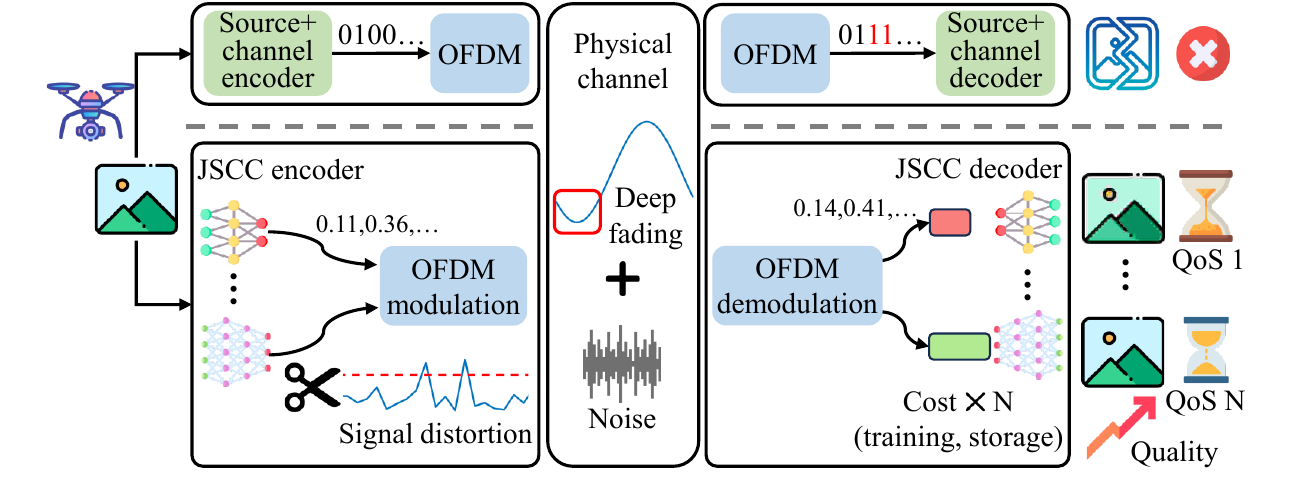}}
    \vspace{-0.4em}
    \caption{Comparison between conventional (top) and DeepJSCC (bottom), along with key challenges in prototyping DeepJSCC.}
    \label{fig_semcom}
\end{figure}

However, most existing works rely on a critical assumption that \textit{the encoded content is transmitted reliably over the wireless channel}~\cite{du2020server, li21sec, xu2011muzi}. In practice, wireless channels are prone to disturbances, leading to errors in the transmitted data, especially under low signal-to-noise ratio (SNR) conditions, as shown in Fig.~\ref{fig_semcom}. In such scenarios, conventional approaches like forward error correction (FEC)~\cite{mackay2005fountain} and retransmission mechanisms often fail to meet stringent latency requirements. This limitation underscores the need for more efficient and noise-resilient transmission schemes tailored to the challenges of wireless environments. 
\textit{Deep joint source-channel coding} (DeepJSCC) has emerged as a promising technique to overcome these challenges~\cite{gunduz2022beyond, luo2022semantic, bourtsoulatze2019deepjscc, gunduz2024joint,dai2022nonlinear}, and is widely regarded as a key enabler for enhancing both transmission efficiency and robustness in next-generation wireless systems. 
Unlike the conventional separation-based approach, as depicted in Fig.~\ref{fig_semcom}, a DeepJSCC system trains the entire ``encoding-decoding'' process end-to-end, with the wireless channel incorporated as an integral part of the model. This holistic approach not only improves transmission efficiency but also enhances robustness against channel-induced noise.

To fulfill this goal, numerous 
DeepJSCC schemes have been proposed to enable efficient and robust transmission of images~\cite{kurka2020deepjscc_f, huang2022toward, xu2021wireless, tang2024contrastive, erdemir2023generative}, speech~\cite{han2022semanticspeech, weng2021semanticspeech}, and video~\cite{wang2022wirelessvideo, tung2022deepwive} over wireless channels. While these schemes have demonstrated superior performance compared to traditional codecs in simulations, usually considering additive white Gaussian noise (AWGN) channel, the scenarios explored in these studies remain largely idealized.
Some efforts have extended 
DeepJSCC to more practical environments, including fading channels and broadband channels, to better reflect real-world wireless conditions~\cite{yang2022jsccofdm, wu2024jsccmimo, mu2022heterogeneous, chi2024capacity, liu2024road}.
However, these studies are still confined to numerical simulations, which may not fully capture the complexities of real-world deployment. 
To fill this gap, we intend to prototype 
DeepJSCC over orthogonal frequency division multiplexing (OFDM) systems, as OFDM serves as the cornerstone of modern wireless technologies (LTE, 5G) and future 6G networks to enable their high throughput and robustness~\cite{li2006orthogonal}.
Our goal is to move beyond theoretical analysis and explore the practical challenges and opportunities associated with deploying 
DeepJSCC in real-world systems.

Building an OFDM-based DeepJSCC system presents several key challenges, as illustrated in Fig.~\ref{fig_semcom}. 
First, unlike traditional coding schemes, DL-based JSCC encoders generate highly correlated feature sequences. 
This violates a fundamental assumption of OFDM, namely that symbols mapped to different subcarriers should be statistically independent~\cite{bolcskei2003impact, park2000papr}, which can significantly degrade performance. 
Second, the strong correlation among these generated symbols also leads to a high peak-to-average power ratio (PAPR)~\cite{shao2022semanticofdm}. More critically, the continuous-valued and unconstrained nature of the semantic feature further exacerbates this issue, leading to nonlinear distortion.  Third, most existing encoders and decoders are trained with a fixed compression ratio~\cite{bourtsoulatze2019deepjscc, xu2021wireless}, while real-world applications require adaptive compression to meet varying latency and quality of service (QoS) constraints. Achieving adaptability in these systems typically requires multiple encoder-decoder pairs, which impose significant computational and storage overhead. Several studies have explored adaptive compression; however, they typically rely on additional side information (e.g., SNR estimates or entropy-related signals)~\cite{yang2022deep, dai2022nonlinear, zhang2023predictive, bian2023deepjscc, yang2024swinjscc}. Conveying such information requires either CSI feedback or an auxiliary digital transmission link, which significantly complicates system implementation.
Finally, existing DeepJSCC systems for video transmission predominantly focus on file transfer, where the entire video is transmitted before playback. These works mainly focus on improving reconstruction performance, while largely overlooking the optimization of computational complexity and latency~\cite{wang2022wirelessvideo, tung2022deepwive, du2025object, zhang2026bi}. In contrast, real-time streaming,  critical for practical deployment,
remains largely underexplored, even in simulation studies. 

In response to these challenges, we propose \name, which, to the best of our knowledge, is the first real-time noise-resilient DeepJSCC prototype designed for efficient and robust multimedia transmission. \name addresses the aforementioned issues through several key innovations. 
To address the first and second challenges, we develop a mapping method from the task-relevant features to the data symbol and a cross-subcarrier precoding method to mitigate inter-symbol correlation across subcarriers and reduce PAPR, thereby enhancing compatibility with OFDM.
Furthermore, we propose a progressive coding strategy that prioritizes the transmission of important features while a tailored training mechanism is incorporated to ensure feature importance decreases gradually. 
Finally, we implement \name on a software-defined radio (SDR) platform,  supporting real-time image transmission and video streaming, 
and further conduct extensive experiments to evaluate \name's performance.
In summary, our main contributions are as follows:
\begin{itemize}
\item We prototype \name, the first OFDM-based real-time noise-resilient 
DeepJSCC system for efficient and robust multimedia transmission.
\item We design a feature-to-symbol mapping method and a cross-subcarrier precoding method to ensure symbol decorrelation and reduce PAPR, enhancing the compatibility of 
DeepJSCC with OFDM.
\item We develop a progressive coding strategy that transmits the important features first, enabling adaptive compression to meet varying transmission latency and QoS constraints.
\item We implement \name on an SDR platform for real-time image transmission and video streaming and validate its performance through extensive experiments.
\end{itemize}

The remainder of this paper is organized as follows. 
Section~\ref{sec:background} provides the background and motivation behind \name. Section~\ref{sec:design} details the system design of \name, including its key components. Sections~\ref{sec:setup} and~\ref{sec:results} present the experiment setup and performance evaluation results, respectively. 
Section~\ref{sec:related_work} briefly discusses related works in the field and discusses the future work.
Section~\ref{sec:conclusion} concludes the paper.

\section{Background and Motivation} \label{sec:background}

In this section, we first introduce the fundamentals of 
DeepJSCC and OFDM, followed by an analysis of why directly integrating them leads to poor performance in practical deployments.

\subsection{DeepJSCC and OFDM}

\begin{figure}[t]
    \setlength{\abovecaptionskip}{6pt}
    \centering
    \includegraphics[width=0.98\linewidth]{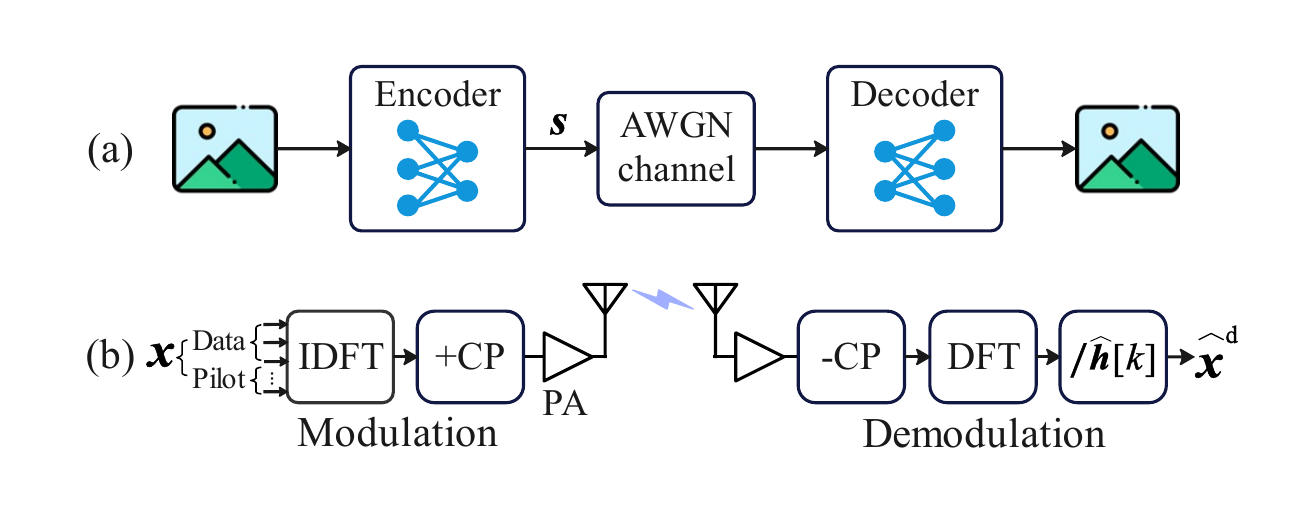}
    \vspace{-0.5em}
    \caption{(a) DeepJSCC workflow and (b) OFDM transmission process.}
    \label{fig:sem_ofdm}
    \vspace{-1.6em}
\end{figure}

\begin{figure*}[t]
    \centering
    \setlength{\abovecaptionskip}{6pt}
    \begin{subfigure}[b]{0.32\linewidth}
        \centering
        \includegraphics[width=0.95\linewidth]{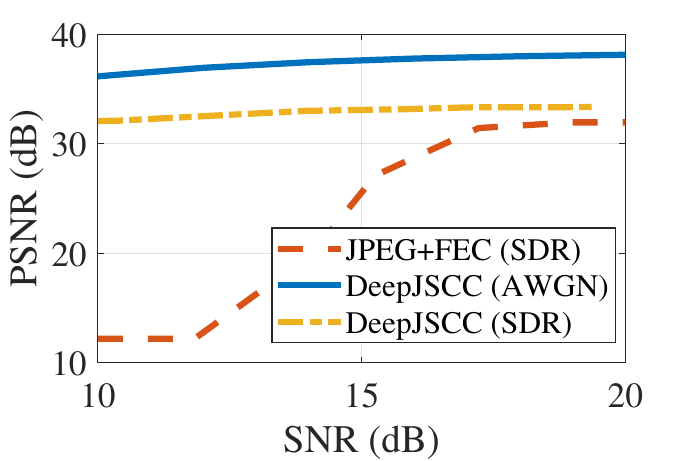}
        \vspace{-0.4em}
        \caption{Performance.}
        \label{fig:performance_awgn}
        \includegraphics[width=0.95\linewidth]{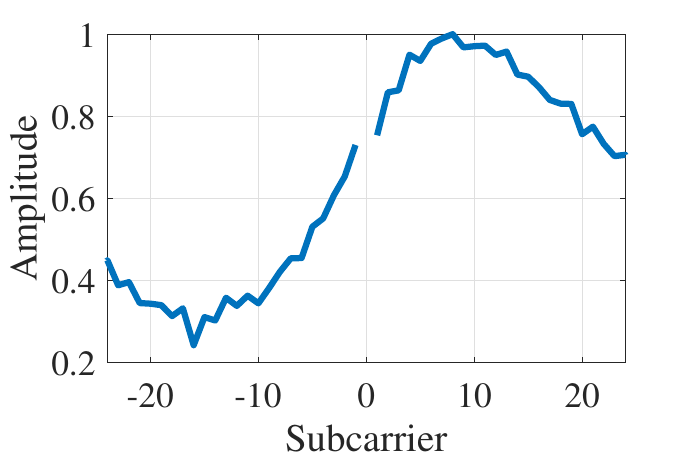}
        \vspace{-0.4em}
        \caption{Fading.}
        \label{fig:channel}
    \end{subfigure}
    \hfill
    \begin{subfigure}[b]{0.32\linewidth}
        \centering
        \includegraphics[width=0.95\linewidth]{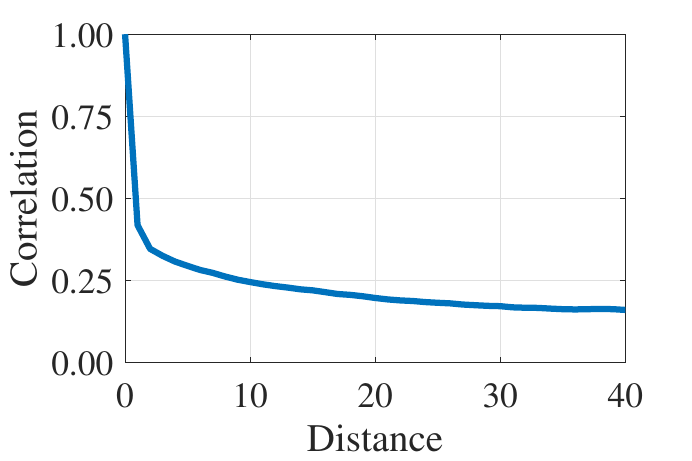} 
        \vspace{-0.4em}
        \caption{Correlation.}
        \label{fig:corr}
        \includegraphics[width=0.95\linewidth]{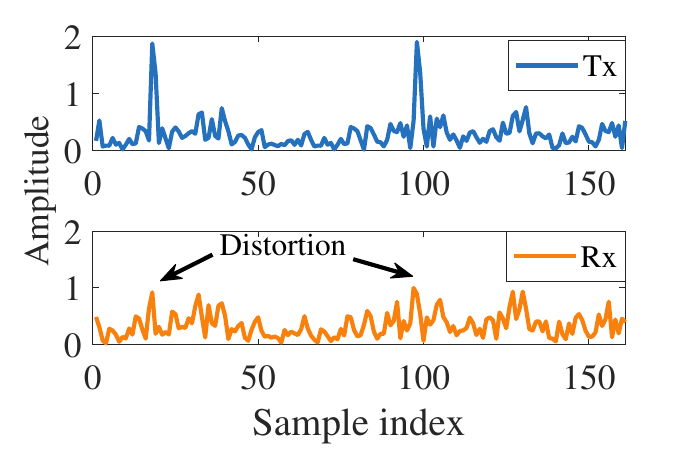}
        \vspace{-0.4em}
        \caption{Signal.}
        \label{fig:txrx_signal}
    \end{subfigure}
    \hfill
    \begin{subfigure}[b]{0.32\linewidth}
        \centering
        \includegraphics[width=0.95\linewidth]{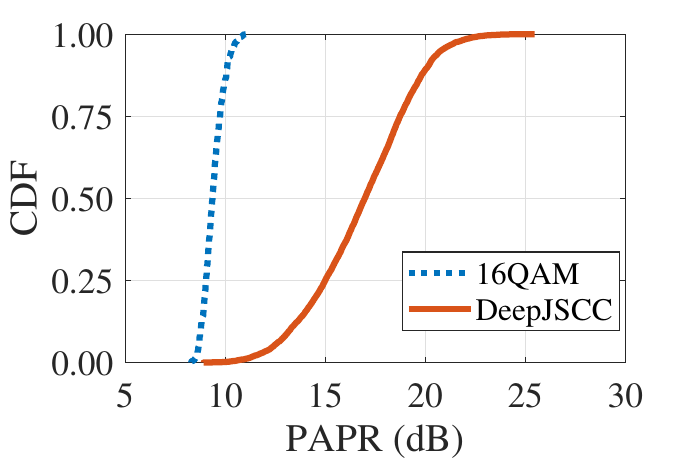}
        \vspace{-0.4em}
        \caption{PAPR.}
        \label{fig:PAPR}
        \includegraphics[width=0.95\linewidth]{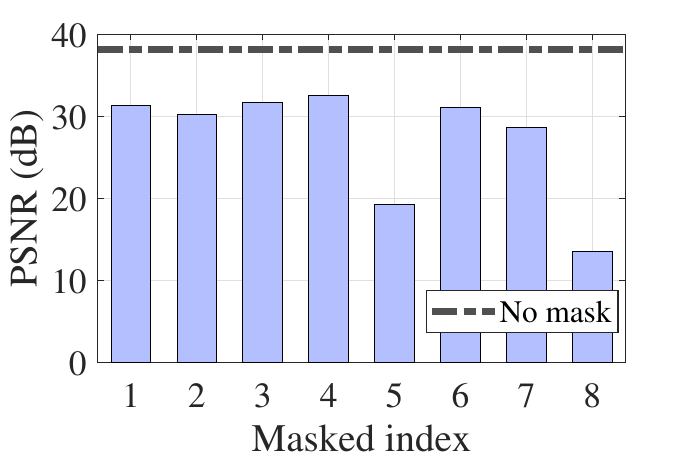}
        \vspace{-0.4em}
        \caption{Masked performance.}
        \label{fig:performance_after_mask}
    \end{subfigure}
    \vspace{-0.2em}
    \caption{(a) Performance over AWGN channel and an SDR platform, (b) frequency-selective fading,
    (c) correlation between two feature elements with different distances,
    (d) transmit and receive signals, (e) cumulative distribution function (CDF) of PAPR, and (f) performance after masking one channel.}
    \vspace{-1.6em}
\end{figure*}

Unlike traditional source and channel coding, DeepJSCC leverages powerful neural networks to perform end-to-end source and channel coding jointly.
Fig.~\ref{fig:sem_ofdm}(a) illustrates a representative DeepJSCC system, consisting of a JSCC encoder parameterized by $\theta$ and a decoder parameterized by $\phi$. 
The encoder extracts task-relevant features $\boldsymbol{s}$ from source content (image or video). A popular choice for the JSCC encoder is a convolutional neural network architecture~\cite{gunduz2022beyond}, where the extracted feature has a size of $H\times W\times C$, where $H$, $W$, and $C$ denote the height, the width, and the number of feature channels, respectively. 
The JSCC decoder leverages the error correction capabilities of neural networks to reconstruct the content.
The encoder and decoder are jointly trained following an end-to-end learning manner, where the channel is incorporated into training to enhance noise resilience.
As shown in Fig.~\ref{fig:performance_awgn}, DeepJSCC achieves a high peak signal-to-noise ratio (PSNR) even at low SNR for image transmission, effectively mitigating the \textit{cliff effect} observed in the conventional scheme with FEC~\cite{bourtsoulatze2019deepjscc}.

To facilitate practical deployment, integrating DeepJSCC with OFDM is essential,
as OFDM has become the foundation of mainstream systems such as Wi-Fi and 4G/5G. Fig.~\ref{fig:sem_ofdm}(b) illustrates the OFDM transmission process, where $K$ denotes the number of subcarriers. The modulation process includes inverse discrete Fourier transform (IDFT) and cyclic prefix (CP) insertion. For an OFDM symbol 
consisting of $K^{\mathrm{d}}$ data symbols $\boldsymbol{x}^{\mathrm{d}}\in\mathbb{C}^{K^{\mathrm{d}}}$ and $(K-K^{\mathrm{d}})$ data pilot (used for phase tracking and calibration~\cite{terry2002ofdm}), the generated OFDM symbol after IDFT is 
\begin{equation}
    \boldsymbol{y} = \boldsymbol{F}^H \boldsymbol{x}^{\mathrm{d}} + \boldsymbol{y}^{\mathrm{p}}, \label{eq:y}
\end{equation}
where $\boldsymbol{F}\in\mathbb{C}^{K^{\mathrm{d}}\times K}$ denotes the 
truncated DFT matrix associated with the data symbols
and $\boldsymbol{y}^{\mathrm{p}}$ corresponds to the data pilot.
To combat inter-symbol interference, the last $L$ entries of $\boldsymbol{y}$ are appended to its beginning, forming the CP-extended signal, which is then transmitted through a power amplifier (PA).
After propagating through the wireless channel, the received signal undergoes OFDM demodulation, which reverses the modulation process.
The received data symbol is denoted as $\hat{\boldsymbol{z}}^{\mathrm{d}}\in\mathbb{C}^{K^{\mathrm{d}}}$, and 
$\hat{{\boldsymbol{z}}}^{\mathrm{d}}[{k}] = \boldsymbol{h}[k] \boldsymbol{x}^{\mathrm{d}}[{k}] +\boldsymbol{n}[k]$, where $\boldsymbol{h}[k]$ is channel gain at the $k$-th subcarrier and $\boldsymbol{n}[k]$ is additive Gaussian noise. To recover $\boldsymbol{x}^{\mathrm{d}}$, the receiver utilizes the estimated channel $\hat{\boldsymbol{h}}[k]$ from preamble pilots to mitigate channel effects: 
\begin{equation}
    \hat{\boldsymbol{x}}^{\mathrm{d}}[{k}] =   \left(\boldsymbol{h}[k] \boldsymbol{x}^{\mathrm{d}}[{k}] +\boldsymbol{n}[k]\right)/{\hat{\boldsymbol{h}}[k]}.
\end{equation}

\subsection{Direct Integration} \label{sec:direct_integration}

The most straightforward approach to integrating DeepJSCC with OFDM is to directly use the continuous-valued features of the JSCC encoder as the input to the OFDM system. In this scheme, features are sequentially transmitted across different feature channels, with elements alternately mapped to the real and imaginary parts of the transmission symbols to maximize spectral efficiency.
Fig.~\ref{fig:performance_awgn} presents the PSNR of this approach on an SDR platform for image transmission. While this scheme achieves a slight performance improvement over the conventional methods, its real-world performance on SDR falls significantly short of the ideal case (AWGN channel). This discrepancy highlights that the system fails to fully harness the potential of DeepJSCC in practical deployment. The primary limitations can be attributed to two fundamental factors.

First, in OFDM, symbols across different subcarriers are assumed to be statistically independent to ensure that deep fades (see Fig.~\ref{fig:channel}) do not simultaneously erase redundant information across multiple subcarriers. Otherwise, decoding accuracy would be severely degraded. 
However, as shown in Fig.~\ref{fig:corr}, the elements in the feature $\boldsymbol{s}$ generated by the JSCC encoder exhibit high correlation, with a correlation around 0.15 even at a separation of 40~elements. This correlation arises from the convolutional structure in the JSCC encoder, where adjacent elements in the sequence originate from neighboring regions of the original image. 
Such strong correlation contradicts the independence assumption of OFDM, increasing the risk of losing entire region information when deep fading occurs and thereby complicating reliable reconstruction at the receiver.

Second, the inherent correlation within the feature also contributes to an elevated PAPR. Given the limited dynamic range of the transmitter’s PA, high peaks in $\boldsymbol{y}$ undergoes nonlinear distortion, as shown in Fig.~\ref{fig:txrx_signal}, making it difficult for the receiver to accurately recover the symbols. Furthermore, the unconstrained nature of feature values generated by the JSCC encoder exacerbates this issue. As shown in Fig.~\ref{fig:PAPR}, compared to OFDM with independently and randomly modulated 16QAM symbols, the direct integration of 
DeepJSCC and OFDM significantly increases the PAPR, leading to a deterioration in the performance.

Since both issues stem from the conflict between the feature's inherent properties of the JSCC encoder and the fundamental assumption of OFDM, we propose two intermediary structures to mitigate these challenges: a feature-to-symbol mapping method (Section~\ref{sec:mapping}) and a cross-subcarrier precoding method (Section~\ref{sec:precoding}). These structures serve as a bridge for aligning the characteristics of 
JSCC encoding with the requirement of OFDM, thereby improving overall system performance.

\subsection{Feasibility Study towards Flexibility} \label{sec:flexibility}

Users typically adjust their transmission requirements based on available bandwidth to prevent excessive delays. For example, adapting the video bitrate ensures smooth real-time playback. However, most existing 
JSCC encoders generally produce fixed-length outputs~\cite{bourtsoulatze2019deepjscc,xu2021wireless}, limiting their flexibility to accommodate varying transmission conditions. To support variable bitrate control, existing approaches require training and storing an individual encoder-decoder pair for each bitrate level, which incurs significant overhead~\cite{kurka2021bandwidth}. Notably, the channel dimension of the output features encodes different aspects of the source content~\cite{zhang2023predictive}. This means that even when only a subset of channels is transmitted, the original data can still be reconstructed, albeit with reduced fidelity.
To verify this, we sequentially mask individual channels and assess the resulted reconstruction performance, as illustrated in Fig.~\ref{fig:performance_after_mask}. The results reveal that masking different channels leads to varying reconstruction quality, indicating that certain channels carry more critical information than others. When the available bandwidth is reduced, we can only transmit the critical channels for meeting latency requirements. 
However, how to pick out the critical channels and further improve the performance using them remains a challenge. Several works enable adaptive transmission through rate adaptation modules; however, they typically rely on additional side information, such as SNR estimates or entropy-related signals, which complicates system implementation~\cite{yang2022deep, dai2022nonlinear, zhang2023predictive, bian2023deepjscc, yang2024swinjscc}. The successive refinement framework in~\cite{kurka2021bandwidth} avoids such side information, but its layer partitioning is relatively coarse, leading to performance degradation.
To address this, Section~\ref{sec:training} introduces a novel progressive coding training strategy that prioritizes encoding essential information in earlier channels. It enables adaptive transmission, where the transmitter can dynamically decide whether to send the later channels based on real-time constraints, thereby meeting user requirements.

\section{The design of \name} \label{sec:design}

\begin{figure}[t]
    \setlength{\abovecaptionskip}{6pt}
    \centerline{\includegraphics[width=0.98\linewidth, trim=5 8 5 8, clip]{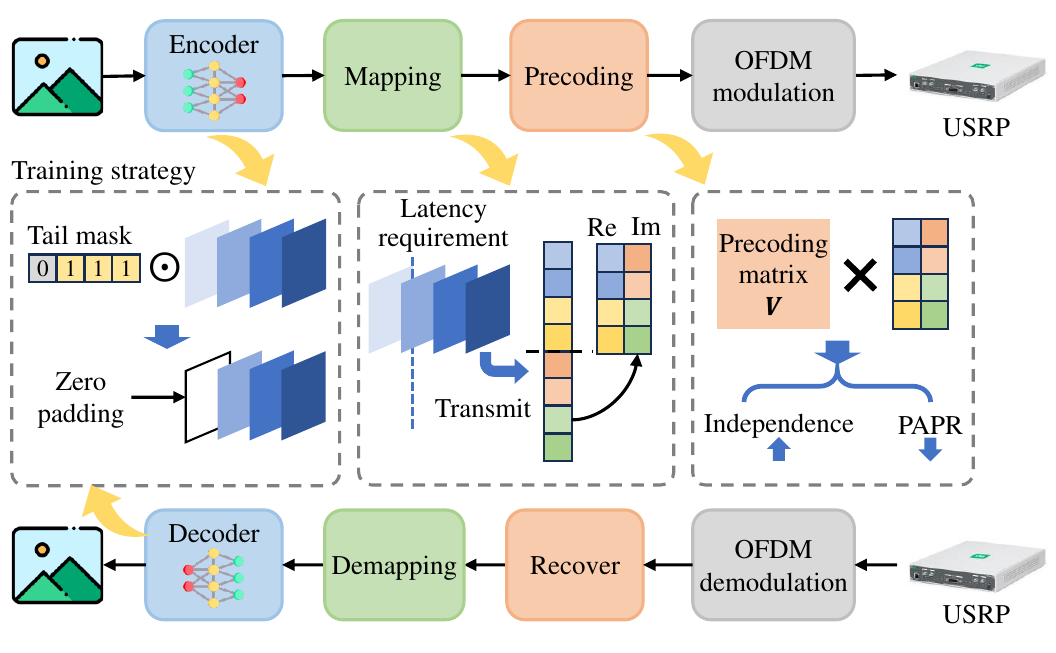}}
    \vspace{-0.5em}
    \caption{{Overview of \name.}}
    \label{fig_overview}
    \vspace{-1.6em}
\end{figure}

Our \name is specifically designed to achieve efficient and noise-resilient real-time multimedia over OFDM. As illustrated in Fig.~\ref{fig_overview}, \name consists of two major components:
\begin{itemize}
\item Feature-to-symbol mapping (Section~\ref{sec:mapping}): It incorporates a flexible progressive coding strategy adapted to varying latency requirements, and a mapping method from feature to symbol.
\item Cross-subcarrier precoding method (Section~\ref{sec:precoding}): It performs symbol decorrelation and PAPR reduction via a tailored precoding matrix.

\end{itemize}
At the receiver, after OFDM demodulation, the signal undergoes inverse precoding and feature-to-symbol inversion before being processed by the 
JSCC decoder to accomplish the target task (e.g., image and video reconstruction).
Furthermore, to support high-performance real-time multimedia transmission, we also develop the following fundamental components:
\begin{itemize}
    \item Joint training strategy (Section~\ref{sec:training}). It ensures that critical information is concentrated in the earlier channels. Additionally, it optimizes both PAPR reduction and compression-reconstruction performance via a loss function.
    \item Real-time streaming framework (Section~\ref{sec:streaming}). It implements a dual-process streaming framework to support low-latency video streaming.
\end{itemize}




\subsection{Feature-to-Symbol Mapping} \label{sec:mapping}

\begin{figure}[t]
    \centering
    \setlength{\abovecaptionskip}{6pt}
    \begin{subfigure}[b]{0.6\linewidth}
        \centering
        \includegraphics[width=0.98\linewidth]{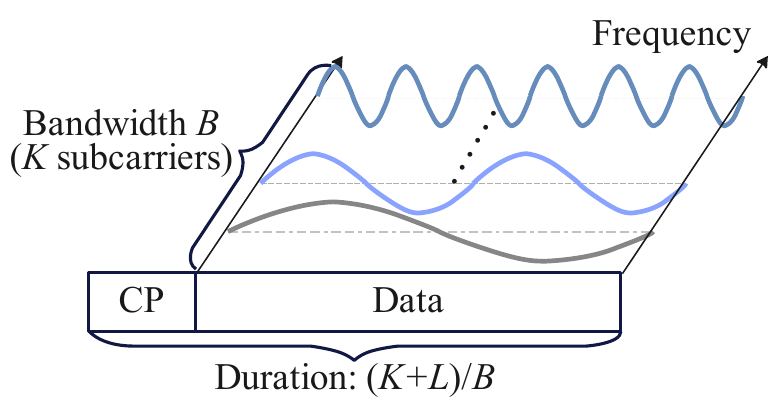} 
        \vspace{-0.3em}
        \caption{OFDM symbol.}
        \label{fig:ofdm_symbol}
    \end{subfigure}
    \begin{subfigure}[b]{0.38\linewidth}
        \centering
        \includegraphics[width=0.975\linewidth]{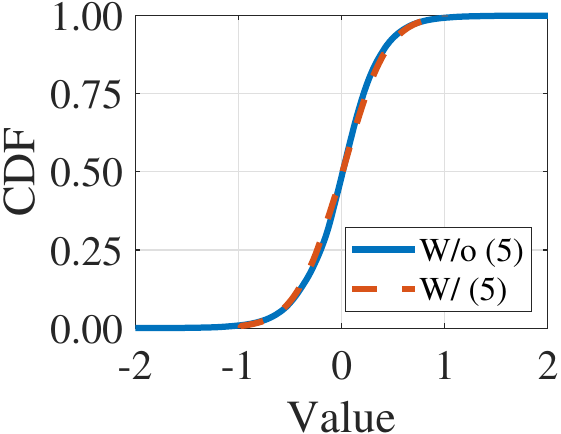}
        \vspace{-0.3em}
        \caption{CDF.}
        \label{fig:cdf_output}
    \end{subfigure}
    \vspace{-0.2em}
    \caption{(a) OFDM symbol structure and (b) value distribution of the encoder output features with/without the proposed activation function.}
    \vspace{-1em}
\end{figure}

In a wireless communication system, users have varying bandwidths, yet they all expect real-time multimedia transmission. To meet this requirement, the amount of transmitted data must be adjusted dynamically.
Specifically, as shown in Fig.~\ref{fig:ofdm_symbol}, given the number of subcarriers $K$, the length of CP $L$, and the user's allocated bandwidth $B$, the duration of one OFDM symbol is $(K+L)/B$. Among the $K$ subcarriers, $K^{\mathrm{d}}$ subcarriers are allocated for data transmission.
Consequently, the achievable symbol rate is ${K^{\mathrm{d}} B}/{(K+L)}$. Unlike conventional coding and modulation schemes, 
DeepJSCC uses the 
JSCC encoder's output directly as the real and imaginary parts of the data symbols. When the required transmission latency is $T^{\max}$, the maximum transmittable feature length is
\begin{equation}
    N = {2K^{\mathrm{d}} B ({T^{\max}}-T^{\mathrm{p}})}/{(K+L)},
\end{equation}
where $T^{\mathrm{p}}$ is the latency occupied by the preamble pilots, and the factor of 2 accounts for both real and imaginary components carrying separate feature elements.
According to the channel masking analysis in Section~\ref{sec:flexibility}, we can drop certain feature channels to meet the real-time requirement $T^{\max}$, and the number of retained feature channels is 
\begin{equation}
    C^{\mathrm{T}} = \left \lfloor \frac{2K^{\mathrm{d}} B { (T^{\max}-T^{\mathrm{p}}) }}{(K+L) HW}\right\rfloor,
\end{equation}
where $\lfloor\cdot \rfloor$ represents the floor operation. Thanks to the training strategy proposed in Section~\ref{sec:training}, feature channels are ranked by importance, from highest to lowest. This implies that earlier channels carry more critical information. Therefore, given a selected number of channels $C^{\mathrm{T}}$, only the first $C^{\mathrm{T}}$ channels are retained for transmission.


Due to the high PAPR issue, $\boldsymbol{s}$ remains unsuitable for direct transmission, as the encoder's output is typically unconstrained. As shown in Fig.~\ref{fig:cdf_output}, the encoder output approximately follows a Gaussian distribution, leading to the occasional occurrence of extreme values. Such outliers induce nonlinear distortion in power amplifiers, thereby degrading overall system performance.
This issue primarily stems from the absence of a suitable activation function. Among commonly used activation functions, ReLU does not impose an upper bound of the output, while Sigmoid and Tanh enforce range constraints but suffer from vanishing gradients at their boundaries~\cite{hanin2018neural}. As a result, none of these functions are well-suited for OFDM-based DeepJSCC.
To address this issue, we utilize a modified activation function that restricts the output range to $\gamma*$[-1,1], as 
\begin{equation}
    f(x) = \min(\max(x,-\gamma),\gamma). 
\end{equation}
By applying it, the encoder output is effectively limited to a predefined range. As illustrated in Fig.~\ref{fig:cdf_output}, where $\gamma$ is set to 1, under the same variance, the value range of the encoder’s output features is significantly reduced, preventing extreme values and thereby mitigating PAPR. Furthermore, this activation function avoids the vanishing gradient issue, facilitating stable joint training of the encoder and decoder. To balance reconstruction performance and PAPR reduction, we set $\gamma=1.5$ in the following experiments.

\begin{figure}[t]
    \centering
    \setlength{\abovecaptionskip}{6pt}
    \begin{subfigure}[b]{0.4\linewidth}
        \centering
        \includegraphics[width=0.85\linewidth]{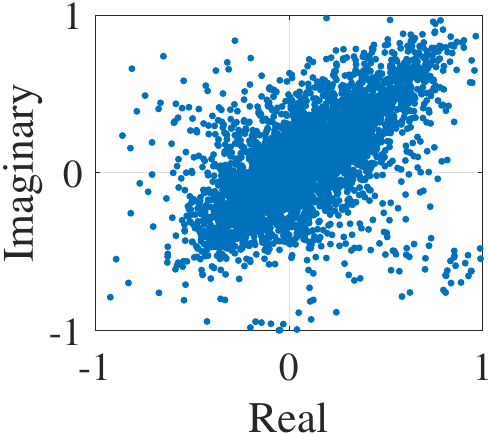} 
        \vspace{-0.3em}
        \caption{Direct.}
        \label{fig:x_mk}
    \end{subfigure}
    \begin{subfigure}[b]{0.4\linewidth}
        \centering
        \includegraphics[width=0.85\linewidth]{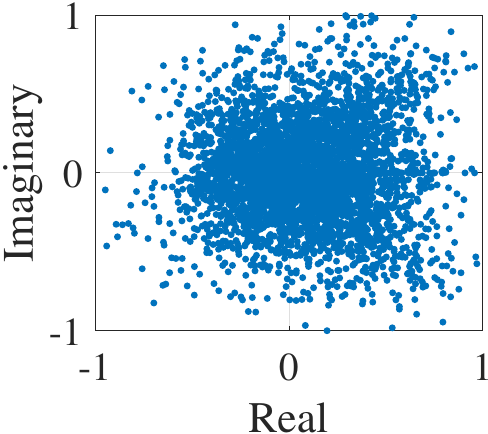}
        \vspace{-0.3em}
        \caption{Proposed.}
        \label{fig:x_mk2}
    \end{subfigure}
    \vspace{-0.2em}
    \caption{The distribution of $\boldsymbol{x}^{\mathrm{d}}$ using (a) directly mapping and (b) the proposed method.}
    \vspace{-1em}
\end{figure}

We now transform the feature into the data symbol $\boldsymbol{x}^{\mathrm{d}}$. First, we segment the feature into multiple subsequences $\boldsymbol{s}$, each of length $2K^{\mathrm{d}}$, and then map them into complex-valued sequences of length $K^{\mathrm{d}}$. However, directly mapping elements in the features alternately to the real and imaginary parts of the data symbol would lead to a low spectral efficiency due to the high correlation between adjacent elements (see Fig.~\ref{fig:corr}). As shown in Fig.~\ref{fig:x_mk}, the resulting $\boldsymbol{x}$ tends to cluster along the line Re=Im in the complex plane. Consequently, the mapped symbols fail to fully exploit the available signal space, resulting in poor spectral efficiency. 
To address this problem, we propose a modified mapping strategy. Specifically, the first half of $\boldsymbol{s}$ is assigned to the real component, while the second half is assigned to the imaginary component, with an alternating sign pattern, yielding $\boldsymbol{x}^{\mathrm{d}}$:
\begin{equation}
    \boldsymbol{x}^{\mathrm{d}}[k] =\left\{
    \begin{array}{ll}
        \!\boldsymbol{s}[k]+j\boldsymbol{s}[k+K^{\mathrm{d}}], & \text{$k$ is odd}, \\
        \!\boldsymbol{s}[k]-j\boldsymbol{s}[k+K^{\mathrm{d}}], & \text{$k$ is even}.
    \end{array}
    \right.
\end{equation}
As shown in Fig.~\ref{fig:x_mk2}, the resulting $\boldsymbol{x}^{\mathrm{d}}$ exhibits a more uniform distribution across the complex plane, effectively improving spectral efficiency.

\subsection{Cross-subcarrier Precoding} \label{sec:precoding}

To address the impact of correlation and reduce PAPR, we introduce a cross-subcarrier precoding matrix, denoted as $\boldsymbol{V}\in\mathbb{C}^{K^{\mathrm{d}}\times K^{\mathrm{d}}}$, where the precoded symbol is given by $\boldsymbol{x}^{\mathrm{t}} = \boldsymbol{V}\boldsymbol{x}^{\mathrm{d}}$.
By appropriately designing $\boldsymbol{V}$, we aim to reduce symbol correlation across subcarriers, conforming to OFDM's fundamental assumption, while also shaping the transmit signal to prevent high PAPR. 
To this end, we first analyze the symbol correlation across subcarriers and the PAPR of the OFDM symbol, followed by an optimization-based design of $\boldsymbol{V}$.

Ensuring symbol independence across subcarriers helps counteract deep fading (see Fig.~\ref{fig:channel}). Ideally, transmitted symbols should be independent across all subcarriers to prevent correlated deep fades from erasing similar information. However, this requirement is overly stringent. In reality, fading correlation decreases with increasing frequency separation, meaning that strict independence is unnecessary for widely spaced subcarriers, as they do not experience similar fading conditions.
To this end, we take subcarrier fading correlation into account and aim to minimize symbol correlation at the receiver.

Specifically, the received symbol at the $k$-th subcarrier is $\boldsymbol{h}[k] \boldsymbol{x}^{\mathrm{t}}[{k}]$ when the noise is ignored, and the correlation between $\boldsymbol{h}[k_1] \boldsymbol{x}^{\mathrm{t}}[{k_1}]$ and $\boldsymbol{h}[k_2] \boldsymbol{x}^{\mathrm{t}}[{k_2}]$ can be expressed as 
\begin{align}
    \boldsymbol{\rho}[k_1,k_2] 
    &=\frac{\mathbb{E}\{(\boldsymbol{h}[k_1] \boldsymbol{x}^{\mathrm{t}}[{k_1}])(\boldsymbol{h}[k_2] \boldsymbol{x}^{\mathrm{t}}[{k_2}]  )\}}{\mathrm{Var}(\boldsymbol{h}[k] \boldsymbol{x}^{\mathrm{t}}[{k}])} \nonumber \\
    & \stackrel{\mathrm{(\star)}}{=} 
    \frac{\mathbb{E}\{ \boldsymbol{h}[k_1]\boldsymbol{h}[k_2] \} \mathbb{E}\{ \boldsymbol{x}^{\mathrm{t}}[{k_1}]\boldsymbol{x}^{\mathrm{t}}[{k_2}])\}}{\mathrm{Var}(\boldsymbol{h}[k]) \mathrm{Var}(\boldsymbol{x}^{\mathrm{t}}[k]) }\nonumber \\
    & = \boldsymbol{\rho}^{{h}}[k_1,k_2] \boldsymbol{\rho}^{{x,\mathrm{t}}}[k_1,k_2],
\end{align}
where $(\star)$ holds due to the independence between $\boldsymbol{x}^{\mathrm{t}}[k]$ and $\boldsymbol{h}[k]$, $\boldsymbol{\rho}^{{h}}$ and $\boldsymbol{\rho}^{{x,\mathrm{t}}}$ represent the covariance matrices for $\boldsymbol{h}$ and $\boldsymbol{x}^{\mathrm{t}}$, respectively, and $\mathbb{E}\{\cdot\}$ is the operation of expectation. In the above expression, $\boldsymbol{\rho}^{{x,\mathrm{t}}}$ can be further given as: $\boldsymbol{\rho}^{{x,\mathrm{t}}} = \boldsymbol{V} \boldsymbol{\rho}^{{x,\mathrm{d}}} \boldsymbol{V}^H$, with $\boldsymbol{\rho}^{{x,\mathrm{d}}}$ being the covariance matrix for $\boldsymbol{x}^{\mathrm{d}}$. Note that $\boldsymbol{\rho}^{{x,\mathrm{d}}}$ can be estimated over the training set of the 
JSCC encoder and decoder. For $\boldsymbol{\rho}^{{h}}$, it can be assumed that subcarriers within the coherence bandwidth are fully correlated, while those beyond the coherence bandwidth are considered uncorrelated~\cite{tse2005fundamentals}.\footnote{The coherence bandwidth depends on the environment, and typical values can be adopted following the existing 4G/5G standards~\cite{3GPP_38.901_R14}.}
Let $K^{\mathrm{c}}$ denote the number of subcarriers within a coherence bandwidth, and $\boldsymbol{\rho}^{{h}}$ is a banded matrix with a bandwidth of $K^{\mathrm{c}}$, where all nonzero elements are equal to 1.

For the PAPR issue, according to Eqn.~\eqref{eq:y}, the OFDM symbol $\boldsymbol{y}$ is expressed as: $\boldsymbol{y} = \boldsymbol{F}^H \boldsymbol{V} \boldsymbol{x}^{\mathrm{d}} + \boldsymbol{y}^{\mathrm{p}}$.
Since designing a distinct precoding matrix for each OFDM symbol is infeasible due to the overhead of transmitting additional recovery information, we instead minimize the expected power at each sampling point, formulated as:
\begin{align}
    \boldsymbol{p}^{{y}} [k]  &= \mathbb{E}\{( \boldsymbol{f}_{k}^H \boldsymbol{V} \boldsymbol{x}^{\mathrm{d}} + \boldsymbol{y}^{\mathrm{p}}[k]) ( \boldsymbol{f}_{k}^H \boldsymbol{V} \boldsymbol{x}^{\mathrm{d}} + \boldsymbol{y}^{\mathrm{p}}[k])^H\} \nonumber\\
    &=p^{\mathrm{t}}\boldsymbol{f}_{k}^H \boldsymbol{\rho}^{{x,\mathrm{t}}} \boldsymbol{f}_{k}  + \boldsymbol{p}^{\mathrm{p}}[k],
\end{align}
where $\boldsymbol{p}^{\mathrm{p}}[k] = \boldsymbol{y}^{\mathrm{p}}[k](\boldsymbol{y}^{\mathrm{p}}[k])^H$ is the transmit power of pilots $\boldsymbol{y}^{\mathrm{p}}$, $p^{\mathrm{t}}$ is the transmit power of each data symbol, and $\boldsymbol{f}_{k}$ is the $k$-th column vector of $\boldsymbol{F}$.

{\begin{algorithm}[b]
    \caption{\small{Optimization for Precoding Matrix}} \label{alg:precoding}
    \KwIn{Covariance matrices $\boldsymbol{\rho}^{h}$ for channel and $\boldsymbol{\rho}^{x,\mathrm{d}}$ for data symbol, and the number of random initializations $N^{\mathrm{r}}$. 
    }
    \KwOut{
    Precoding matrix $\boldsymbol{V}$.
    }
    \vspace{.3em}
    \For{$n=1,\cdots, N^{\mathrm{r}}$}{
    Randomly initialize $\boldsymbol{V}$\; 
    %
    %
    
    \For{$k = 1,\cdots,K^{\mathrm{d}}$}
    {
    obtain $\boldsymbol{v}_k$ by solving the problem with $\boldsymbol{v}_k \boldsymbol{v}_{k'}^H = 0$, $k'< k$ and $||\boldsymbol{v}_k ||^2\le 1$\;
    $\boldsymbol{v}_k \leftarrow  \boldsymbol{v}_k/||\boldsymbol{v}_k|| $\;
    }
    Construct $\boldsymbol{V}_n$ and calculate the corresponding objective function value $O_n$.
    }
    Compare the $N^{\mathrm{r}}$ objective values and select the $\boldsymbol{V}$ corresponding to the minimum one as the output.
\end{algorithm}
}

Given the above analysis, we can formulate the following problem to optimize the precoding matrix:
\begin{equation}
    \!\!\!\min_{\boldsymbol{V}}\!\sum_{k_1,k_2} \boldsymbol{\rho}^{{h}}[k_1,k_2] \boldsymbol{\rho}^{{x,\mathrm{t}}}[k_1,k_2] \!+\! \omega \max_{k}  \boldsymbol{p}^{{y}}[k],~\mathrm{s.t.}~\boldsymbol{V}\boldsymbol{V}^H \!=\! \boldsymbol{I},\! 
\end{equation}
where $\omega$ is a weighting factor balancing PAPR reduction and the unitary constraint on $\boldsymbol{V}$ ensures that the transmit power per each subcarrier remains unchanged. This problem is non-convex due to the non-convex objective function and constraint. To address this problem, we leverage the alternating optimization method~\cite{bezdek2003convergence}. Considering the structure of $\boldsymbol{\rho}^{{h}}[k_1,k_2]$, we optimize each row vector of $\boldsymbol{V}$ sequentially. Specifically, when optimizing the $k$-th row vector, i.e., $\boldsymbol{v}_k$, the objective function is convex because $\sum_{k_1,k_2}\boldsymbol{\rho}^{{h}}[k_1,k_2]$ is a quadratic function of $\boldsymbol{v}_k$ with a positive quadratic term and $\boldsymbol{p}^{{y}}[k]$ follows the same convex property. The unitary constraint is decomposed into two separate constraints: one requires $\boldsymbol{v}_k$ to be orthogonal to the previous $(k-1)$ column vectors, i.e., $\boldsymbol{v}_k \boldsymbol{v}_{k'}^H = 0$, $k'< k$, which is convex, while the other enforces the norm of $\boldsymbol{v}_k$ to be 1, i.e., $||\boldsymbol{v}_k||^2 = 1$, which is non-convex. To handle the non-convex constraint, we first relax the non-convex constraint into a convex one: $||\boldsymbol{v}_k||^2 \le1$, and then optimize $\boldsymbol{v}_k$ using existing convex solvers, such as CVX~\cite{CVX}. 
The resulting $\boldsymbol{v}_k$ is then normalized to satisfy the original constraint.
After $K$ steps, the final precoding matrix $\boldsymbol{V}$ is obtained.
Additionally, to further minimize the objective function, we initialize multiple instances of $\boldsymbol{V}$ randomly and apply the optimization process to each. The solution with the lowest objective function value is selected as the final $\boldsymbol{V}$. The complete optimization procedure is outlined in Algorithm~\ref{alg:precoding}.
Notably, the proposed optimization process is primarily related to the type of transmission task (associated with $\boldsymbol{\rho}^{{x,\mathrm{d}}}$) and the wireless environment (associated with $\boldsymbol{\rho}^{h}$). 
Each subproblem in the proposed precoding design can be formulated as a convex quadratically constrained quadratic programming (QCQP) problem, whose computational complexity is approximately $\mathcal{O}({(K^{\mathrm{d}})}^{3})$. Since the optimization consists of $K^{\mathrm{d}}$ independent subproblems, the computational complexity to obtain a solution is approximately $\mathcal{O}({(K^{\mathrm{d}})}^{4})$. It is worth noting that Algorithm~1 is executed only once during the offline deployment stage to obtain the precoding matrix, after which no further updates are required. Therefore, the associated computational overhead is incurred only once and is negligible compared with the overall neural network training cost. We note that although precoding matrix is trained from specific datasets and channel condition, subsequent extensive experiments demonstrate that the precoding matrix performs well under diverse wireless environments and different datasets, exhibiting its robustness and generality.

\subsection{Joint Training Strategy} \label{sec:training}

\begin{figure}[t]
    \setlength{\abovecaptionskip}{6pt}
    \centering
    \includegraphics[width=0.96\linewidth]{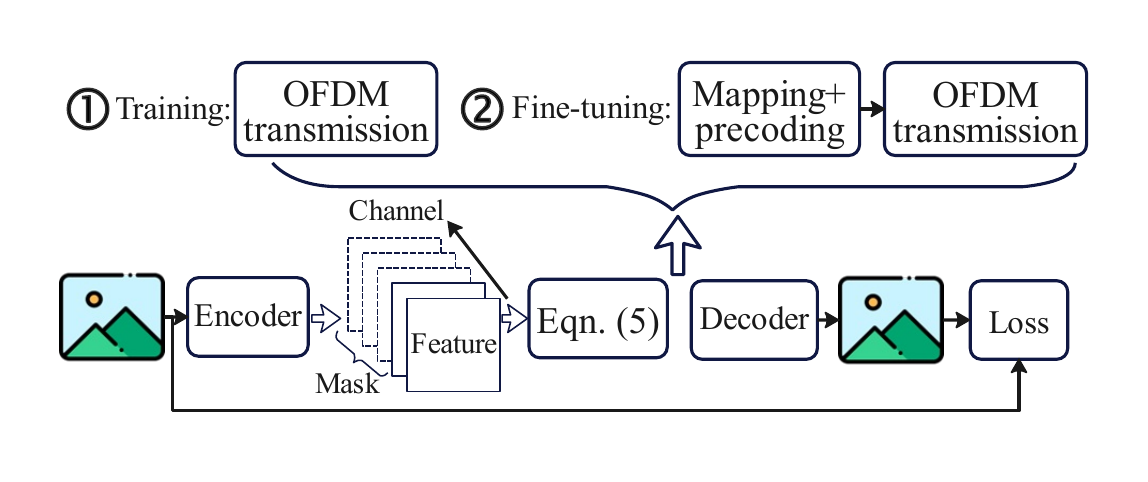}
    \vspace{-0.3em}
    \caption{Two-step training strategy.}
    \label{fig:training}
    \vspace{-1.8em}
\end{figure}

To ensure the effectiveness of DeepJSCC and the designs in Sections~\ref{sec:mapping} and~\ref{sec:precoding} over real-world channels, we propose a two-step training strategy, as shown in Fig.~\ref{fig:training}. The first step is the progressive coding and decoding training over the multipath channel, aiming to concentrate critical information in the earlier channels. This step ensures the system's ability to handle varying latency and QoS requirements.
The second step is fine-tuning to adapt to the proposed mapping and precoding methods.

In the first step, to ensure that critical information is concentrated in the earlier channels, inspired by~\cite{yang2022deep, kurka2021bandwidth}, 
we introduce a random masking strategy. Specifically, at each training iteration, we randomly generate a mask length $C^{\mathrm{m}}$ ($0 \le C^{\mathrm{m}} < C_o$), where $C_o<C$ is the maximum number of removable channels and $C$ is the total number of feature channels). The last $C^{\mathrm{m}}$ feature channels in the encoder output are then discarded and replaced with zeros to maintain dimensional consistency. To guarantee full-performance operation, the zero-mask case is sampled with probability 0.5, while the remaining masking cases share the other 0.5 probability uniformly. This process can be viewed as alternately optimizing multiple channel-budget objectives across iterations~\cite{koike2020stochastic}. Compared with prior works, our method achieves a finer-grained progressive masking scheme than~\cite{kurka2021bandwidth}, while eliminating the dependency on additional side information required by~\cite{yang2022deep}.
For image/video reconstruction tasks, we use mean squared error (MSE) as the loss function, ensuring that the decoder learns to restore the original input with high fidelity.
Through this training approach, the encoder naturally prioritizes encoding important information into earlier channels, while the decoder learns to adapt to such feature distributions, improving overall performance.

Before starting the second step, we first use the trained encoder and decoder from the first step to estimate the covariance matrix of $\boldsymbol{x}^{\mathrm{d}}$, which is subsequently used to calculate the precoding matrix $\boldsymbol{V}$.
Next, we proceed with the second step, where the JSCC encoder and decoder are fine-tuned to adapt to OFDM transmission. The channel state information (CSI) samples for fine-tuning are randomly sampled.
This fine-tuning process incorporates the mapping method and activation function from Section~\ref{sec:mapping}, as well as the OFDM transmission process with the precoding matrix from Section~\ref{sec:precoding}.
Additionally, to mitigate the PAPR issue, we introduce a weighted loss function combining MSE and PAPR~\cite{shao2022semanticofdm}, formulated as:
\begin{equation}
    \mathcal{L} = \alpha \mathrm{MSE}+(1-\alpha) \mathrm{PAPR}.
\end{equation}
In the sequel, we set $\alpha=1e^{-5}$.
This ensures a balance between reconstruction accuracy and PAPR reduction, enhancing both semantic fidelity and transmission efficiency. 

\subsection{Real-Time Streaming} \label{sec:streaming}

Before deploying the 
JSCC encoder and decoder after training, we can further reduce hardware overhead through quantization. Since our feature transmission method resembles analog transmission, the feature SNR is inherently limited by the resolution of the SDR's ADC and DAC. Given that the precision of ADC and DAC is typically limited to 16 bits, we apply 16-bit floating-point quantization to the model (originally 32-bit). This significantly reduces encoding and decoding latency while maintaining near-identical performance.

\begin{figure}[t]
    \setlength{\abovecaptionskip}{6pt}
    \centering
    \includegraphics[width=0.78\linewidth]{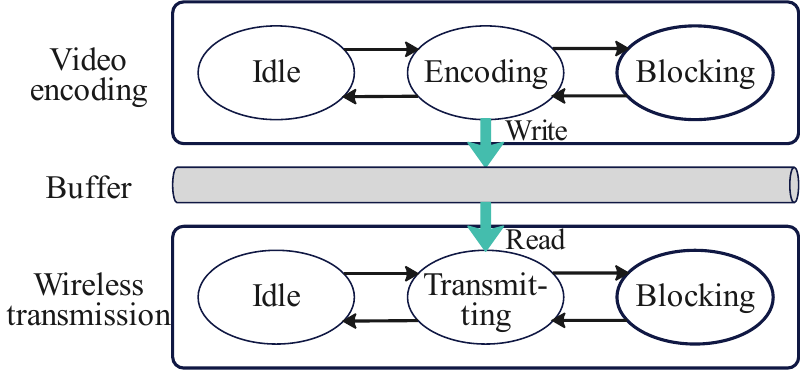}
    \vspace{-0.3em}
    \caption{Dual-process streaming framework at the transmitter.}
    \label{fig:video_stream}
    \vspace{-1.2em}
\end{figure}

For image transmission, we adopt a sequential transmission approach, where the 
JSCC encoder first extracts features, followed by wireless transmission using SDR. At the receiver, the signal is demodulated and decoded to complete the reconstruction task. 
Unlike image transmission, video transmission requires continuous frame transmission to enable real-time streaming. A sequential approach would introduce high latency, as each frame would need to wait for the previous frame to be fully decoded. To address this issue, we employ a dual-process parallel streaming framework, as illustrated in Fig.~\ref{fig:video_stream}. In this framework, one process handles video encoding, while the other manages wireless transmission, with a buffer serving as temporary storage for the encoded features. Specifically, the first process (video encoding) operates in three states: idle, encoding, and blocking. In the idle state, it waits for a new frame to arrive. Once a frame arrives, it enters the encoding state, where the 
JSCC encoder compresses the frame into features and writes them to the buffer. If the buffer is full, the process enters the blocking state, waiting until the buffer becomes available before returning to the encoding state. After writing data, it transitions back to the idle state, ready to process the next incoming frame. In the second process for wireless transmission, there are also three states: idle, transmitting, and blocking. Different from the first process, the transmitting state involves mapping, precoding, OFDM modulation, and signal transmission. The blocking state occurs when the process waits for the channel to become available before continuing to transmit. At the receiver, a similar framework is employed for reconstruction, with implementation details omitted for brevity.



\section{Implementation and Setup} \label{sec:setup}
In this section, we first elaborate on \name's implementation and then introduce the experiment setup.
\subsection{System Implementation} \label{ssec:implementation}

\begin{figure}[t]
    \centering
    \setlength{\abovecaptionskip}{6pt}
    \begin{subfigure}[b]{0.99\linewidth}
        \centering
        \includegraphics[width=0.98\linewidth, trim= 25 40 25 40, clip]{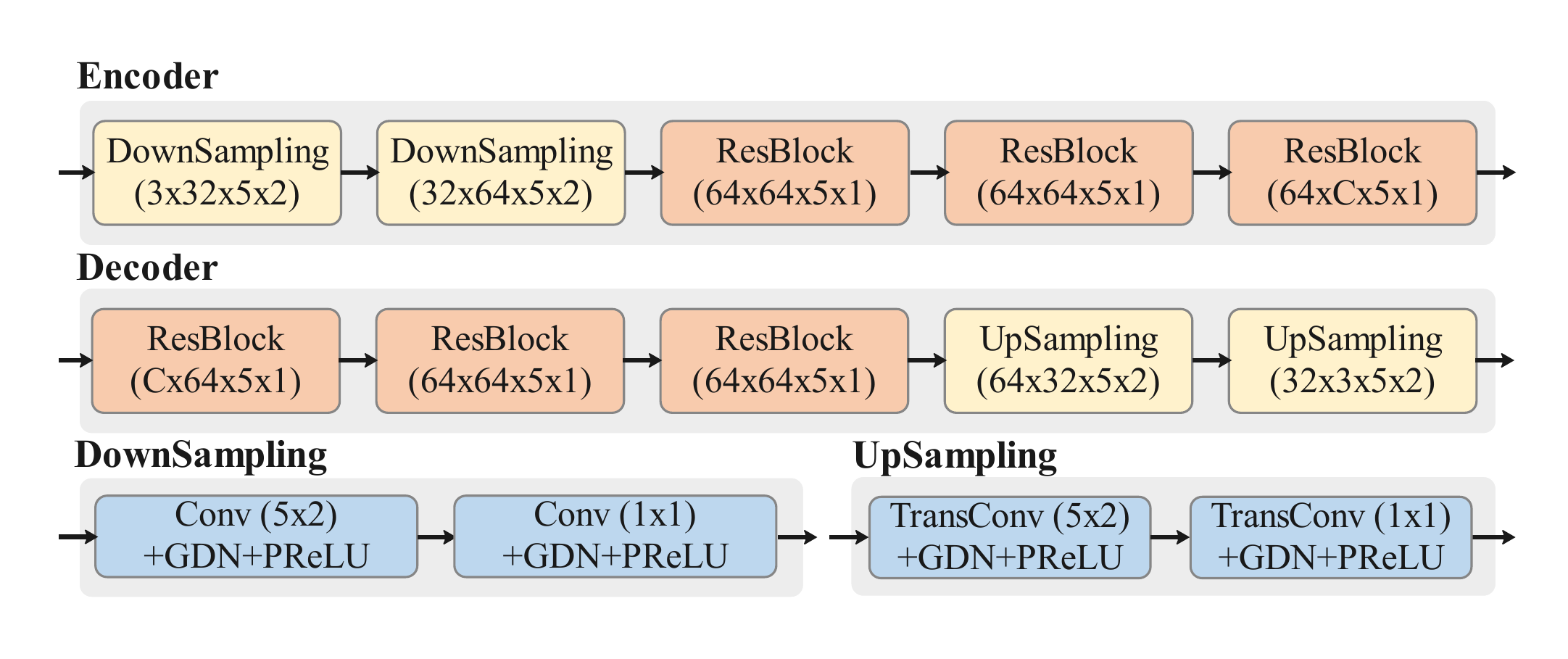}
        \vspace{-0.3em}
        \caption{Encoder and decoder for image transmission.}
        \label{fig:net_image}
    \end{subfigure}
    \begin{subfigure}[b]{0.99\linewidth}
        \centering
        \includegraphics[width=0.98\linewidth]{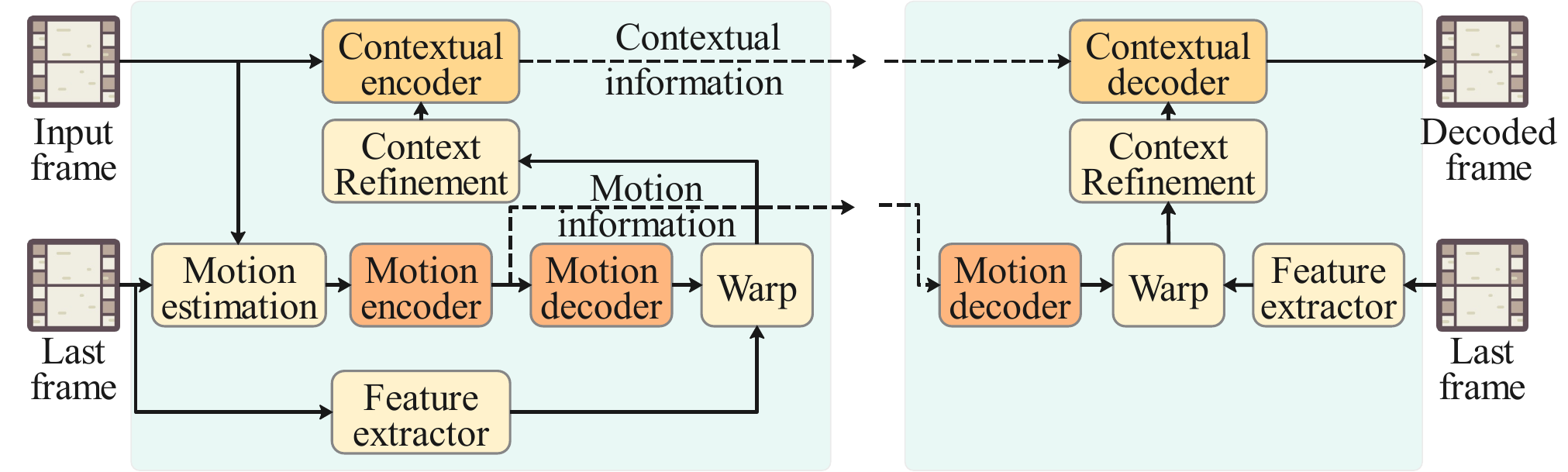}
        \vspace{-0.2em}
        \caption{Encoder and decoder for video transmission.}
        \label{fig:net_video}
    \end{subfigure}
    \vspace{-0.3em}
    \caption{Encoder and decoder for (a) image transmission and (b) video streaming.}
    \vspace{-1.8em}
\end{figure}

\begin{figure}[t]
    \centering
    \setlength{\abovecaptionskip}{6pt}
    \begin{subfigure}[b]{0.45\linewidth}
        \centering
        \includegraphics[width=0.98\linewidth]{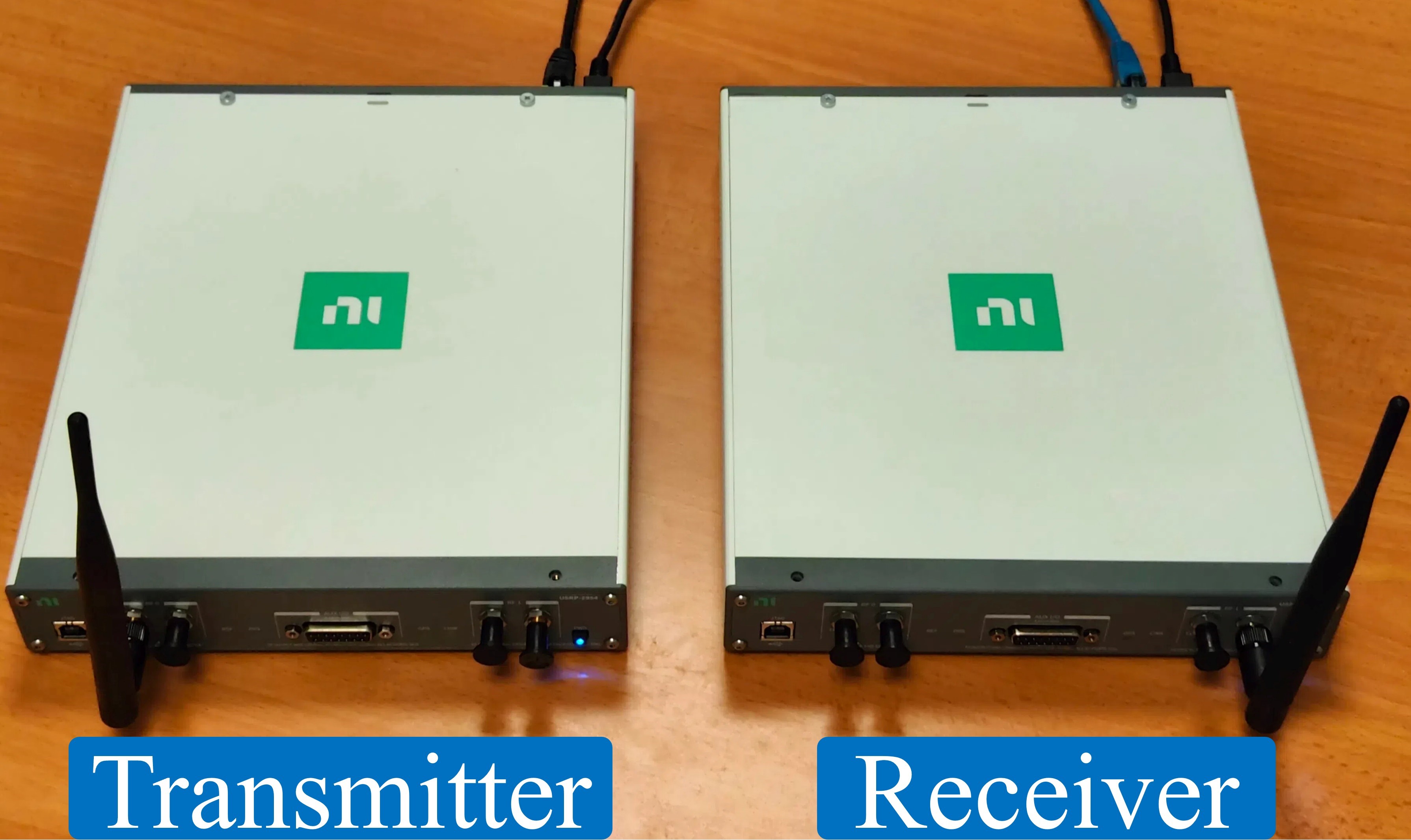}
        \vspace{-0.5em}
        \caption{Hardware.}
        \label{fig:hardware}
    \end{subfigure}
    \begin{subfigure}[b]{0.5\linewidth}
        \centering
        \includegraphics[width=0.98\linewidth]{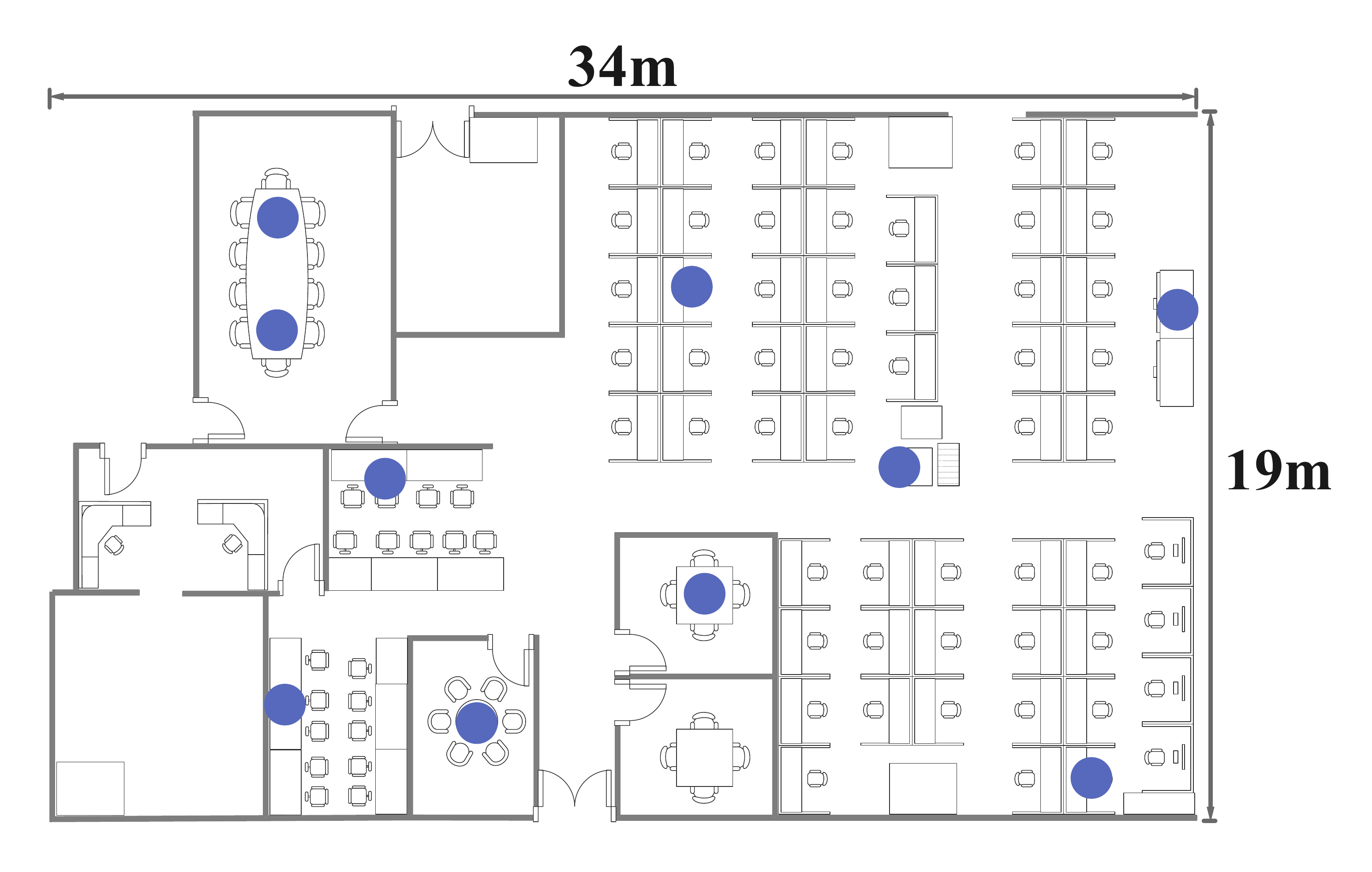}
        \vspace{-0.5em}
        \caption{Experiment layout.}
        \label{fig:experiment_layout}
    \end{subfigure}
    \caption{\name implementation: (a) hardware configurations and (b) experiment layout.}
    \vspace{-1.8em}
\end{figure}

In this paper, we consider image transmission and video streaming as two representative multimedia applications. The architecture and training details of the 
JSCC encoder and decoder are described below.

\textit{Image transmission}. As shown in Fig.~\ref{fig:net_image}, The 
JSCC encoder consists of five convolutional layers, including three residual layers~\cite{he2016deep} and two downsampling layers, which reduce the feature's width and height to 1/4 of the original image. The number of channels is set to 12.
Similarly, the decoder is also built with five convolutional layers, and the downsampling layers are replaced with upsampling layers to ensure that the reconstructed image retains the original dimensions. Both 
JSCC encoder and decoder are implemented in Python 3.8 with PyTorch 1.13 and trained on a Linux server equipped with RTX A6000 GPUs, following the training strategy in Section~\ref{sec:training}. The UCF-101 dataset~\cite{soomro2012ucf101} is used for training and testing, where images are randomly cropped to 256$\times$256 pixels. Adam optimizer is adopted, with a learning rate of $5\times10^{-4}$ and a batch size of 32.

\textit{Video streaming}. For video streaming, we handle the transmission of both keyframes (I-frames) and predicted frames (P-frames). Following \cite{tung2022deepwive}, we set the group-of-pictures (GOP) size as 4. Since I-frames require direct frame transmission, we adopt the image transmission approach with $C$ being 8. For P-frames, instead of transmitting the entire frame, we transmit only the differences relative to the previous frame, thereby reducing the transmission load. To achieve this, we adopt the method proposed in~\cite{li2021deepvideo}. As shown in Fig.~\ref{fig:net_video}, this method extracts and transmits motion features and contextual features sequentially. Both motion and contextual information are downsampled to 1/16 of the original image's height and width, with the number of motion feature channels and contextual feature channels set to 16 and 64, respectively.
The training process follows the same procedure as image transmission, using the UCF-101 dataset with a learning rate of $1\times10^{-4}$ and a batch size of 8.

With the trained encoder and decoder, we implement a \name prototype using two USRP X310 devices~\cite{usrpx310} and two workstations, as shown in Fig.~\ref{fig:hardware}. One pair of USRP and a workstation serve as the transmitter, while the other functions as the receiver. 
We utilize the USRP Hardware Driver (UHD)~\cite{uhd} to manage signal transmission and reception. Both USRP devices operate at a center frequency of 2.4~\!GHz with a bandwidth of 10~\!MHz. Following the general OFDM configurations~\cite{crow1997ieee}, the system employs 64 subcarriers, with 48 allocated for data symbols, 4 for data pilot, and 12 left unused. The default transmission latency constraint is set to 3~\!ms.

\subsection{Experiment Setup}
We conduct experiments in an office environment, as illustrated in Fig.~\ref{fig:experiment_layout}. Ten locations are selected as transmitter and receiver positions to measure the average performance under different SNR conditions in both line-of-sight (LoS) and non-line-of-sight (NLoS) scenarios. The total over-the-air measurement duration exceeds 18 hours. While DeepJSCC is known to perform well under low-SNR conditions, our real-world experiments focus on the 10–20~\!dB range due to practical testbed constraints. First, in over-the-air experiments, noise power is estimated using transmitted zero padding, and the estimation becomes unreliable at very low SNR. Second, packet detection becomes significantly less reliable as noise increases, leading to high packet loss and frequent retransmissions.

\begin{figure*}[t]
    \centering
    \setlength{\abovecaptionskip}{6pt}
    \begin{subfigure}[b]{0.32\linewidth}
        \centering
        \includegraphics[width=1\linewidth]{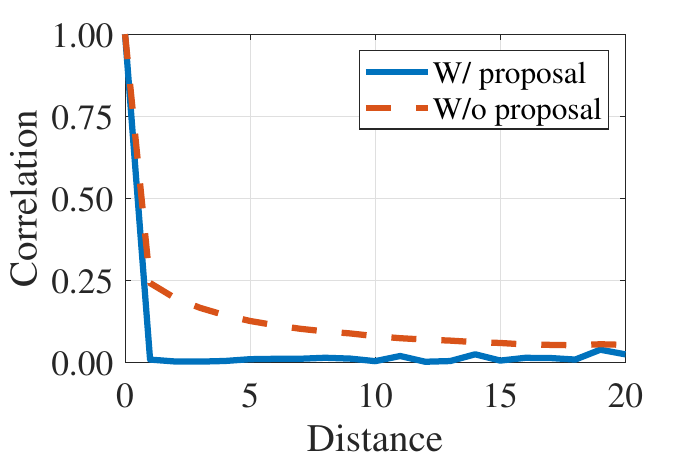}
        \vspace{-1.5em}
        \caption{Symbol correlation.}
        \label{fig:symbol_corr}
        \includegraphics[width=1\linewidth]{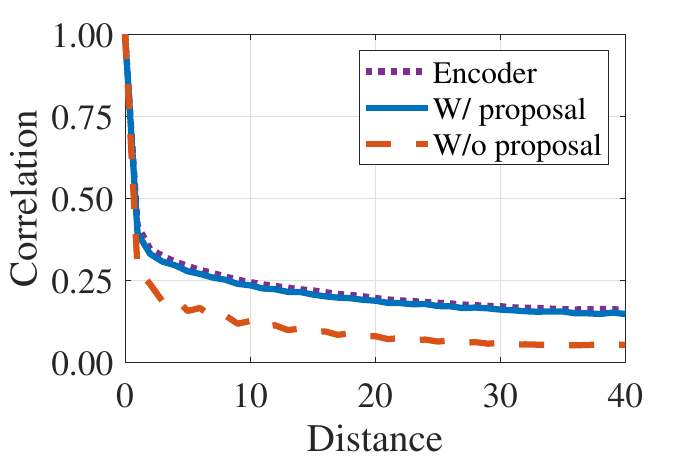}
        \vspace{-1.5em}
        \caption{Feature correlation.}
        \label{fig:feature_corr}
    \end{subfigure}
    \hfill
    \begin{subfigure}[b]{0.32\linewidth}
        \centering
        \includegraphics[width=1\linewidth]{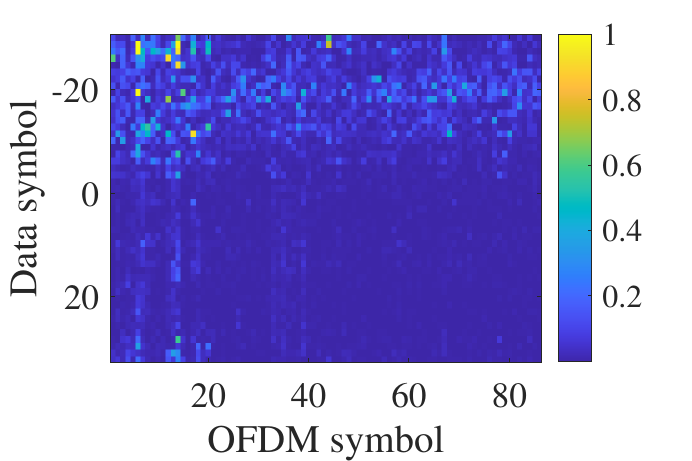}
        \vspace{-1.5em}
        \caption{W/o proposal.}
        \label{fig:mse_baseline}
        \includegraphics[width=1\linewidth]{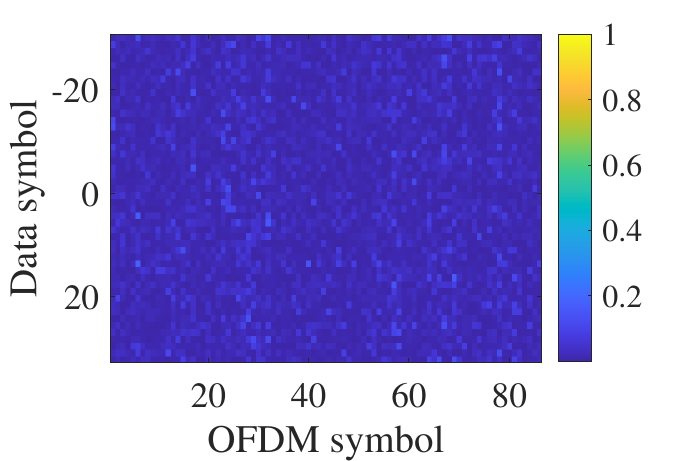}
        \vspace{-1.5em}
        \caption{W/ proposal.}
        \label{fig:mse_propose}
    \end{subfigure}
    \hfill
    \begin{subfigure}[b]{0.32\linewidth}
        \centering
        \includegraphics[width=1\linewidth]{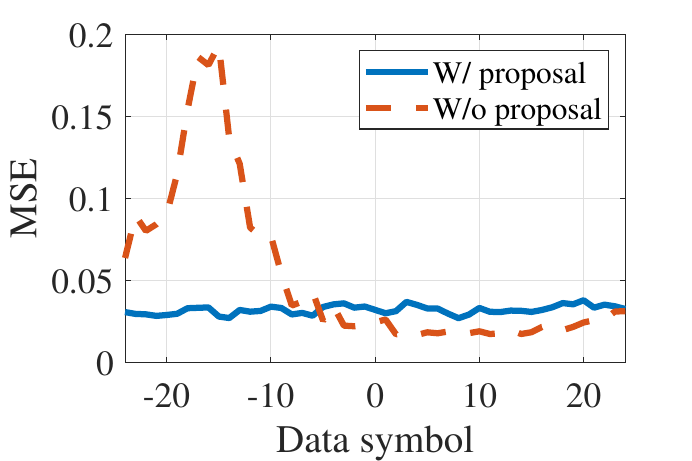}
        \vspace{-1.5em}
        \caption{MSE.}
        \label{fig:microbench_mse_carrier}
        \includegraphics[width=1\linewidth]{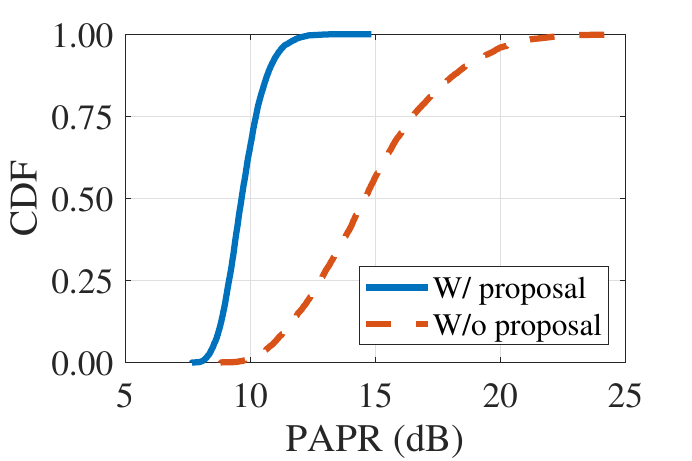}
        \vspace{-1.5em}
        \caption{PAPR.}
        \label{fig:microbench_papr}
    \end{subfigure}
    \caption{Performance of the proposed mapping and precoding methods: (a) symbol correlation, (b) correlation between two feature elements, (c) and (d) squared error across different subcarriers and OFDM symbols, (e) MSE across data symbols, and (f) PAPR.}
    \vspace{-1.5em}
\end{figure*}

For image transmission, we consider three baselines:
i) direct deployment of existing DeepJSCC~\cite{bourtsoulatze2019deepjscc}, where the same neural network mentioned in Section~\ref{ssec:implementation} is adopted without our proposed modifications and further trained with OFDM; 
ii) JSCC-OFDM proposed in~\cite{yang2022jsccofdm}, which is specially designed for OFDM systems. The authors introduce a clipping module to reduce the PAPR. Following the original paper, we set the clipping ratio to $\rho=1.4$, which was reported to provide the best balance between PAPR reduction and reconstruction performance. We adopt the modified OFDM+CE+EQ variant reported in the original paper, where the CE/EQ refinement subnetworks are not used. To ensure a fair comparison, we employ the same OFDM receiver pipeline for all schemes by using least squares (LS)-based channel estimation and zero-forcing (ZF) equalization instead of the original minimum mean square error (MMSE)-based receiver. Furthermore, we replace the original encoder and decoder neural networks with the same encoder and decoder architecture adopted by our method and retrain the model using the same dataset. This implementation ensures that any performance differences are attributable to the proposed intermediate-layer system design rather than differences in the receiver algorithms or neural network architecture.
iii) standard image codecs, where JPEG is used for source coding, low-density parity-check (LDPC) code~\cite{gallager1962low} is adopted for channel coding with a block length of 672 and a code rate of 1/2~\cite{tang2024contrastive}, and 16QAM is applied for OFDM transmission. Following~\cite{bourtsoulatze2019deepjscc}, when JPEG decoding fails, the image is reconstructed using the mean pixel value for each channel.


For video transmission, we also consider three baselines: i) direct deployment of existing neural network (where entropy coding module is removed)~\cite{li2021deepvideo}, where the same neural network mentioned in Section~\ref{ssec:implementation} is adopted without our proposal and further trained with OFDM; 
ii) standard video codecs, where H.264 is adopted for source coding, LDPC code is employed for channel coding, and 16-QAM is used for modulation.
iii) advanced video codecs, where H.265 is adopted for source coding, LDPC coding is employed for channel coding, and 16-QAM is used for modulation.

To ensure a fair comparison, all schemes are evaluated under the same transmission latency during the experiment. The compression ratios of JPEG and H.264 are adjusted according to the channel code rate and modulation order to meet the transmission latency requirement. For image transmission, PSNR is employed as the performance metric. For video transmission, we adopt the multiscale structural similarity index measure (MS-SSIM) to evaluate the quality of the reconstructed frames. Following recent work \cite{wang2022wirelessvideo}, we express MS-SSIM in dB using the formula $-10\log(1-\text{MS-SSIM})$.

\section{Evaluation} \label{sec:results}
We start with three micro-benchmark studies to demonstrate
the effectiveness of each component in \name and then present the overall performance.

\begin{figure*}[t]
    \centering
    \setlength{\abovecaptionskip}{6pt}
    \begin{subfigure}[b]{0.32\linewidth}
        \centering
        \includegraphics[width=1\linewidth]{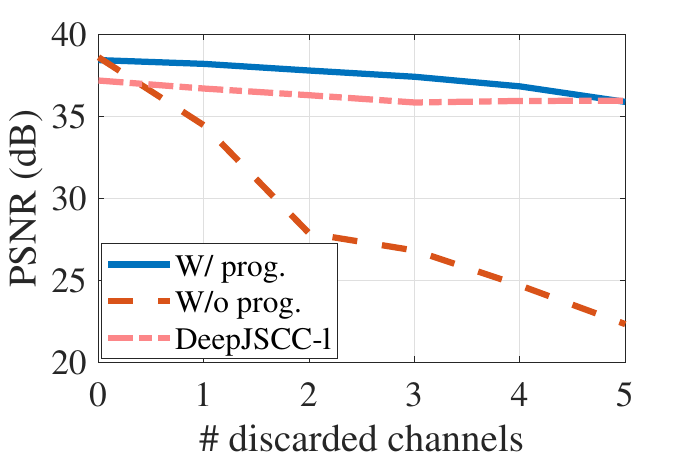}
        \vspace{-1.5em}
        \caption{Image (Progressive coding).}
        \label{fig:microbench_Prog_UCFImage_PSNR}
        \includegraphics[width=1\linewidth]{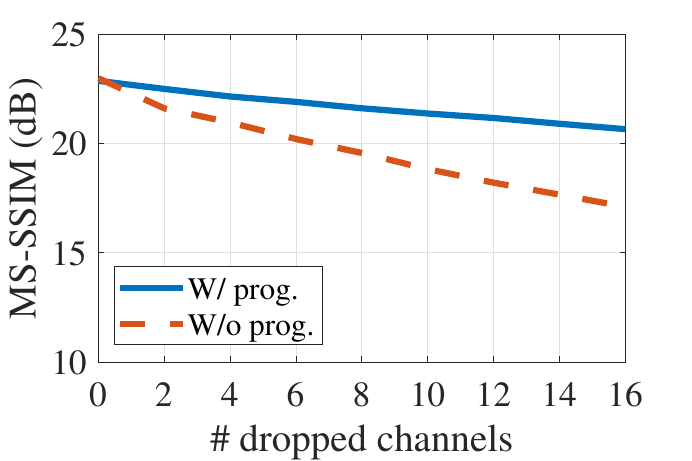}
        \vspace{-1.5em}
        \caption{Video (Progressive coding).}
        \label{fig:microbench_Prog_UCFVideo_}
    \end{subfigure}
    \hfill
    \begin{subfigure}[b]{0.32\linewidth}
        \centering
        \includegraphics[width=1\linewidth]{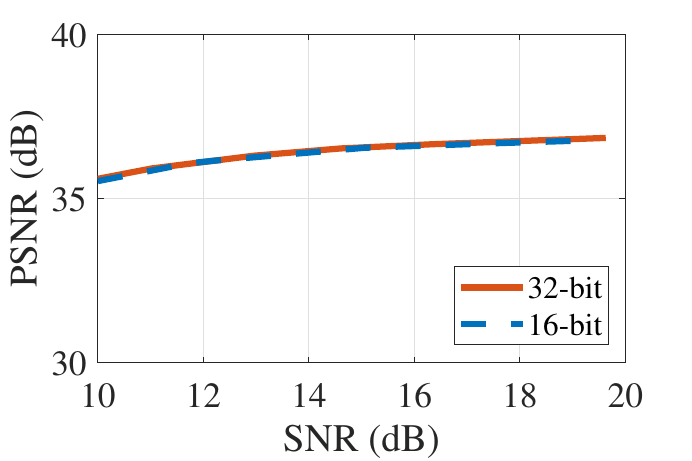}
        \vspace{-1.5em}
        \caption{Image (Quantization).}
        \label{fig:quantization_image}
        \includegraphics[width=1\linewidth]{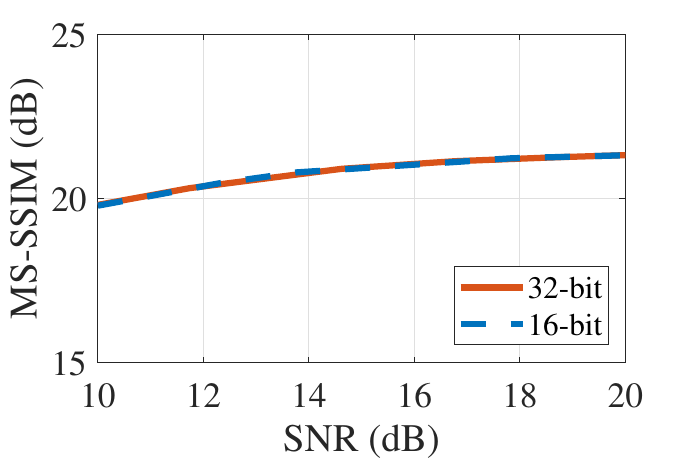}
        \vspace{-1.5em}
        \caption{Video (Quantization).}
        \label{fig:quantization_video}
    \end{subfigure}
    \hfill
    \begin{subfigure}[b]{0.32\linewidth}
        \centering
        \includegraphics[width=1\linewidth]{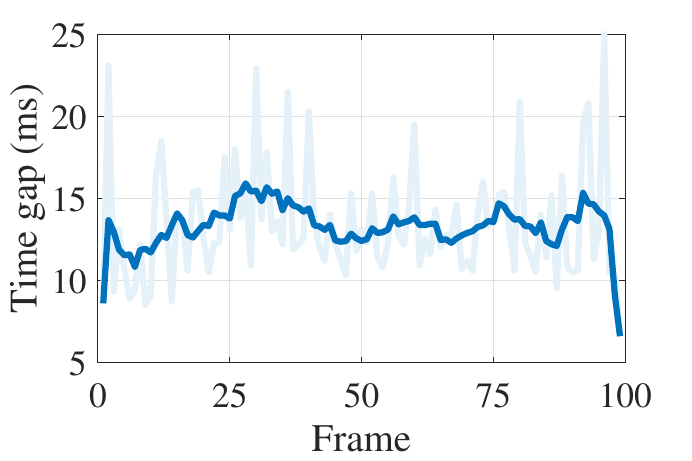}
        \vspace{-1.5em}
        \caption{Time gap.}
        \label{fig:stream_time}
        \includegraphics[width=1\linewidth]{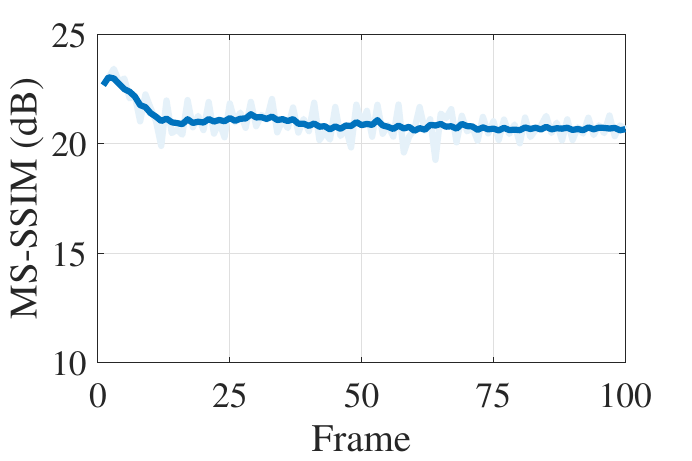}
        \vspace{-1.5em}
        \caption{MS-SSIM.}
        \label{fig:stream_msssim}
    \end{subfigure}
    \caption{Performance of (a) image and (b) video transmission with different number of discarded channels, and performance with/without quantization for (c) image transmission and (d) video streaming, and video streaming: (e) time gap between adjacent frames and (f) MS-SSIM of each frame.}
    \label{fig:microbench_2}
    \vspace{-1.8em}
\end{figure*}

\subsection{Micro-benchmark Studies} \label{mico-benchmark}
\subsubsection{Mapping and precoding}

To evaluate the effectiveness of the feature-to-symbol mapping and precoding methods proposed in Sections~\ref{sec:mapping} and~\ref{sec:precoding}, Fig.~\ref{fig:symbol_corr} presents the symbol correlation between two different subcarriers at the receiver. One can see that, the proposed mapping and precoding methods significantly reduce the correlation compared to the baseline without these techniques, meeting the requirements of OFDM. 
Additionally, Fig.~\ref{fig:feature_corr} illustrates the correlation between elements in the demodulated feature at the receiver.
The results show that, compared to the baseline, the feature correlation better aligns with that generated by the JSCC encoder, demonstrating that the characteristic of feature is well preserved after OFDM modulation and wireless transmission. This property enhances information recovery performance at the JSCC decoder.

To further assess the effectiveness of the mapping and precoding methods, we present the squared error of different data symbols across various OFDM symbols, as shown in Figures~\ref{fig:mse_baseline} and~\ref{fig:mse_propose}. Without the proposed methods, each data symbol at the same position within an OFDM symbol corresponds to a fixed subcarrier, making it susceptible to deep fading, as shown in Fig.~\ref{fig:mse_baseline}. The figure reveals that data symbols on certain subcarriers experience high squared error, while some OFDM symbols exhibit consistently high squared error due to nonlinear distortion caused by high PAPR.
In contrast, as depicted in Fig.~\ref{fig:mse_propose}, the proposed methods effectively mitigate squared error fluctuations, leading to a more stable transmission.
Furthermore, Fig.~\ref{fig:microbench_mse_carrier} shows the MSE across different data symbol positions over multiple OFDM symbols, confirming that our method equalizes the error distribution, thereby preventing localized information loss.
Finally, Fig.~\ref{fig:microbench_papr} depicts the CDF of PAPR, demonstrating a substantial reduction in PAPR when applying our method. These results collectively validate the effectiveness of the proposed mapping and precoding methods in mitigating subcarrier correlation, reducing squared error, and improving transmission robustness.

\subsubsection{Progressive coding}
We now validate the effectiveness of the progressive encoding strategy. 
We discard a portion of the later channels in the feature, and then directly feed the remaining features into the decoder for reconstruction after undergoing the channel. 
Fig.~\ref{fig:microbench_Prog_UCFImage_PSNR} illustrates the image transmission performance under different numbers of discarded channels. For comparison, we also present the performance of a baseline scheme without progressive coding. As shown in the figure, although both schemes experience performance degradation as more channels are discarded, our proposed strategy exhibits a significantly slower decline. Even when five channels are discarded, the performance loss of our scheme remains within 3~\!dB, whereas the baseline suffers a 16~\!dB drop. We also compare our method with DeepJSCC-l~\cite{kurka2021bandwidth}, which divides feature channels into a base layer and a refinement layer and trains the system by randomly masking the refinement layer. To ensure a fair comparison, we adopt the same neural network architecture and vary only the training strategy. As shown in the figure, our method achieves superior performance when fewer than five channels are discarded, demonstrating the effectiveness of our fine-grained progressive transmission strategy.

We further evaluate the effectiveness of progressive encoding in video transmission, focusing on P-frames, as shown in Fig.~\ref{fig:microbench_Prog_UCFVideo_}
Here, the number of discarded channels refers to the total number of motion and contextual channels, each contributing equally. The results confirm that our proposal also outperforms the baseline. When 16 channels are discarded, our scheme incurs only a 2.2~\!dB performance loss, whereas the baseline suffers a 5.8~\!dB loss. These results underscore the effectiveness of the progressive encoding strategy in maintaining high QoS. 
In the sequel, the number of motion feature channels and contextual feature channels is set to 8 and 56, respectively.

\subsubsection{Quantization and video streaming}

Before conducting the streaming tests, we first present the results when the encoder and decoder are quantized to 16-bit floating-point precision in Fig.~\ref{fig:quantization_image} and Fig.~\ref{fig:quantization_video}. The figure demonstrates that for both image and video transmission, the quantized encoder and decoder achieve nearly identical performance to the original 32-bit floating-point ones across different SNRs. This can be attributed to the inherent quantization limitations of the USRP's ADC and DAC, which dictate the overall system precision.
Therefore, as long as the model's quantization bit width is not lower than the precision of the ADC and DAC, performance remains virtually unaffected.

We next evaluate the real-time video streaming performance of \name. We select one representative sequence for visualization. The blurred curves represent the frame-wise measurements,  while the solid curves denote the corresponding fitted trends for better visualization. Fig.~\ref{fig:stream_time} depicts the time gap between the completion of consecutive frame decodings at the receiver, where most gaps fall within 20~\!ms. For a 30~FPS video (frame interval of 33~\!ms), this gap satisfies the real-time playback requirement, validating the effectiveness of our dual-process framework. Furthermore, Fig.~\ref{fig:stream_msssim} presents the MS-SSIM of each decoded frame. The results show consistently high MS-SSIM values, providing initial evidence of \name's superior performance. Moreover, we evaluate the system using ten video sequences, each containing 100 frames. The average end-to-end latency is 2050 ms with the proposed dual-processing framework, compared with 2454 ms without it. For the latter case, the encoding, decoding, and transmission stages account for 1084 ms, 817 ms, and 553 ms, respectively. This corresponds to a latency reduction of up to 16.5\%, demonstrating the effectiveness of the proposed framework for low-latency video streaming.
In the next section, we will further evaluate \name's performance.

\begin{figure*}[t]
    \centering
    \setlength{\abovecaptionskip}{6pt}
    \begin{subfigure}[b]{0.32\linewidth}
        \centering
        \includegraphics[width=1\linewidth]{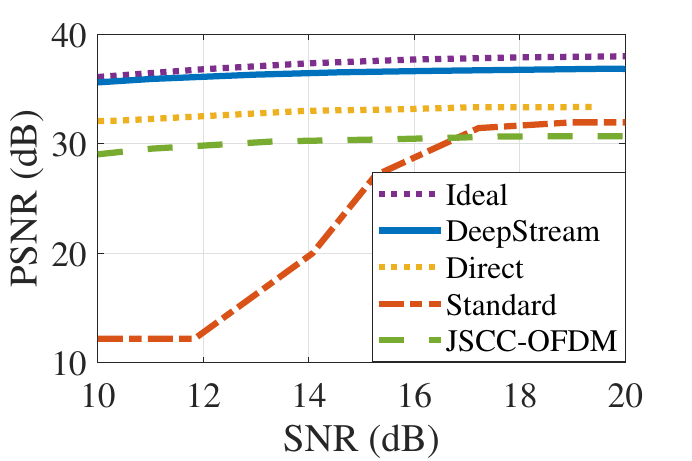}
        \vspace{-1.5em}
        \caption{Image (Overall).}
        \label{fig:end2end_image}
        \includegraphics[width=1\linewidth]{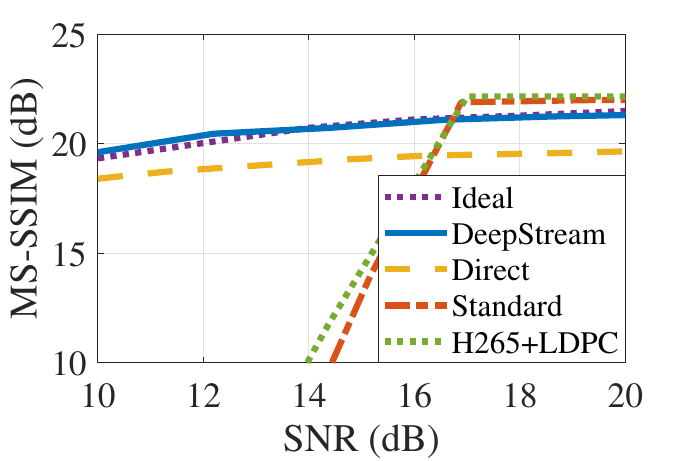}
        \vspace{-1.5em}
        \caption{Video (Overall).}
        \label{fig:end2end_video}
    \end{subfigure}
    \hfill
    \begin{subfigure}[b]{0.32\linewidth}
        \centering
        \includegraphics[width=1\linewidth]{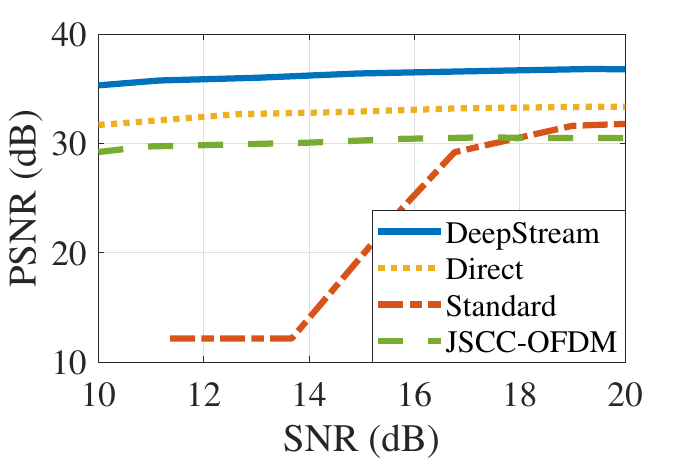}
        \vspace{-1.5em}
        \caption{Image (NLoS).}
        \label{fig:NLoS_image}
        \includegraphics[width=1\linewidth]{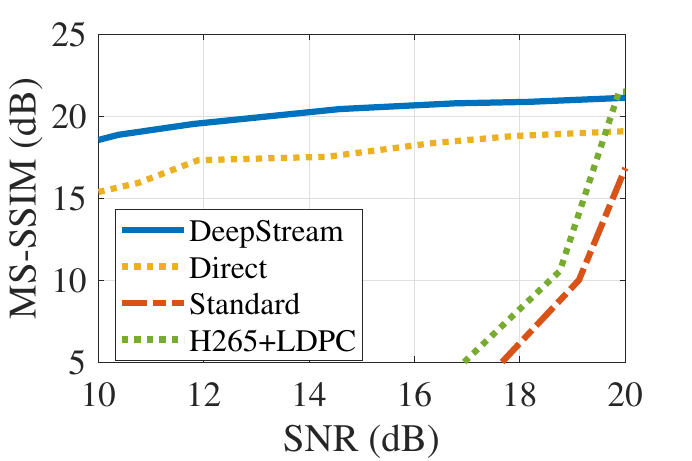}
        \vspace{-1.5em}
        \caption{Video (NLoS).}
        \label{fig:NLoS_video}
    \end{subfigure}
    \hfill
    \begin{subfigure}[b]{0.32\linewidth}
        \centering
        \includegraphics[width=1\linewidth]{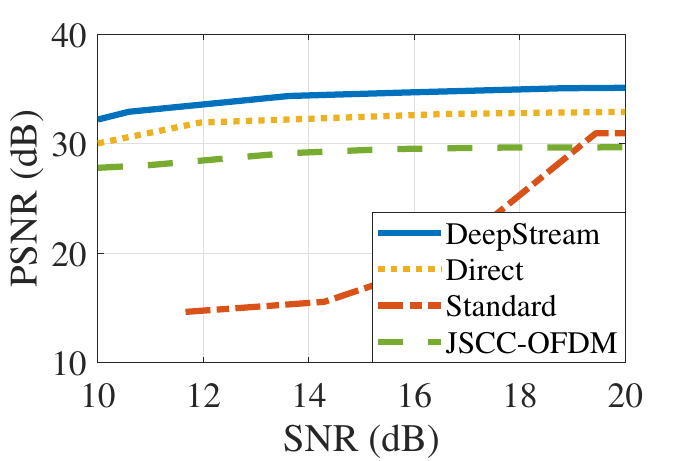}
        \vspace{-1.5em}
        \caption{Image (Generalization).}
        \label{fig:general_image}
        \includegraphics[width=1\linewidth]{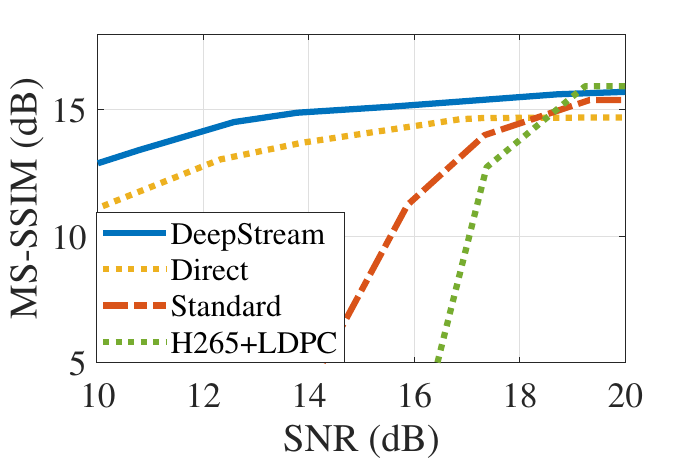}
        \vspace{-1.5em}
        \caption{Video (Generalization).}
        \label{fig:general_video}
    \end{subfigure}
    \caption{Performance under various SNRs for (a) image transmission and (b) video streaming, performance for (c) image transmission and (d) video streaming in the NLoS scenario, and generalization performance for (e) image transmission with Kodak dataset and (f) video streaming with HEVC Class D dataset.}
    \label{fig:exp_1}
    \vspace{-1.8em}
\end{figure*}

\subsection{End-to-end Performance}
\subsubsection{Overall performance}
Fig.~\ref{fig:end2end_image} and Fig.~\ref{fig:end2end_video} present the performance of image transmission and video streaming under different SNR conditions in the LoS scenario.
For image transmission, all schemes experience performance degradation as the SNR decreases. However, the decline is significantly slower for all DeepJSCC-based schemes compared to the standard scheme. Notably, at a lower SNR level, the standard scheme suffer from the cliff effect, whereas DeepJSCC approaches maintain relatively high performance. Moreover, \name demonstrates a significant performance improvement over the direct deployment of DeepJSCC. At an SNR of 10~\!dB,  \name achieves an approximate 3~\!dB performance gain.
To assess whether \name fully exploits the potential of DeepJSCC, we compare its performance against the ideal case under AWGN conditions. The results indicate that \name closely approaches the ideal performance, with a maximum loss of only 1~\!dB. JSCC-OFDM shows significant performance degradation compared with the proposed scheme and even performs worse than the direct deployment baseline. This is because JSCC-OFDM performs clipping directly on the transmitted OFDM waveform. Although the network is trained jointly with the clipping module, the clipping distortion is introduced at the waveform level and directly alters the transmitted signal. Consequently, the performance loss caused by aggressive clipping may outweigh its benefit in PAPR reduction, which is consistent with the observations reported in~\cite{yang2022jsccofdm}. In contrast, our method adopts a fundamentally different strategy. Instead of clipping the transmitted waveform, we only impose a mild bounded activation on the encoder output to suppress extreme feature values. Furthermore, the proposed feature-to-symbol mapping and cross-subcarrier precoding modules further reduce PAPR without introducing additional signal distortion, since both operations are invertible transformations. As a result, the burden of PAPR reduction is shared across multiple modules rather than relying solely on clipping. This enables the generated OFDM signal to exhibit a lower PAPR while preserving the semantic information carried by the learned features, allowing the decoder to maintain stable reconstruction performance.
Similarly, Fig.~\ref{fig:end2end_video} presents the results for video transmission, revealing a comparable trend. Specifically, the standard scheme and H.265+LDPC outperform the DeepJSCC-based schemes at high SNR regimes. However, their performance deteriorates rapidly as the SNR decreases, whereas the DeepJSCC schemes maintain robust performance in low-SNR regimes. Moreover, \name consistently outperforms the direct deployment scheme, thereby validating the effectiveness of the proposed framework.

\begin{figure*}[t]
    \centering
    \setlength{\abovecaptionskip}{6pt}
    \begin{subfigure}[b]{0.32\linewidth}
        \centering
        \includegraphics[width=1\linewidth]{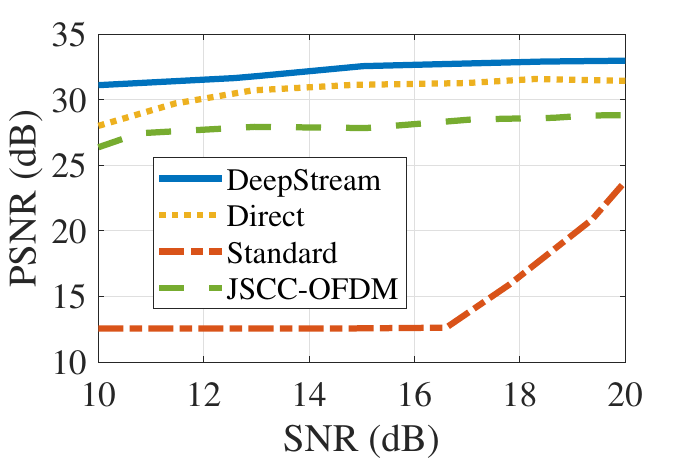}
        \vspace{-1.5em}
        \caption{Image (Generalization 2).}
        \label{fig:general_2_image}
        \includegraphics[width=1\linewidth]{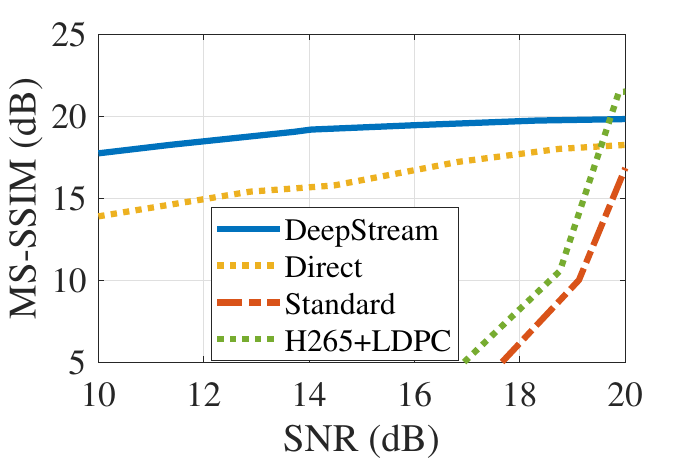}
        \vspace{-1.5em}
        \caption{Video (Generalization 2).}
        \label{fig:general_2_video}
    \end{subfigure}
    \hfill
    \begin{subfigure}[b]{0.32\linewidth}
        \centering
        \includegraphics[width=1\linewidth]{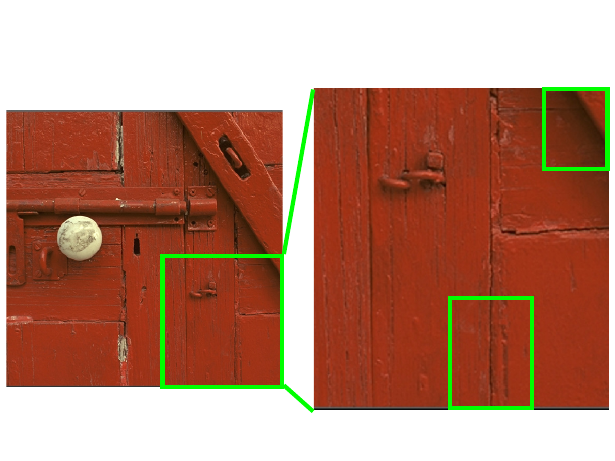}
        \vspace{-1.5em}
        \caption{Original.}
        \label{fig:example_1}
        \includegraphics[width=1\linewidth]{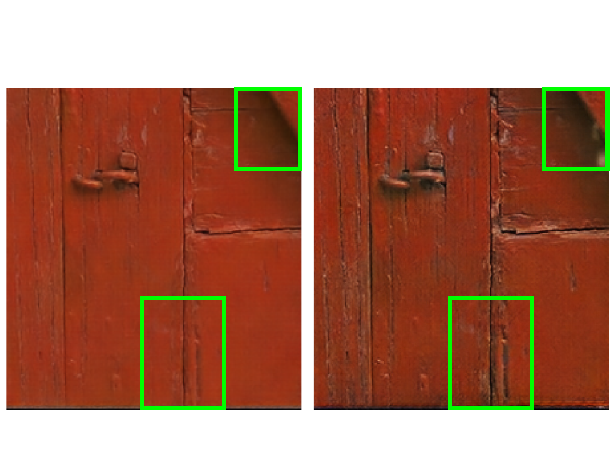}
        \vspace{-1.5em}
        \caption{\name (left) and Direct (right).}
         \label{fig:example_2}
    \end{subfigure}
    \hfill
    \begin{subfigure}[b]{0.32\linewidth}
        \centering
        \includegraphics[width=1\linewidth]{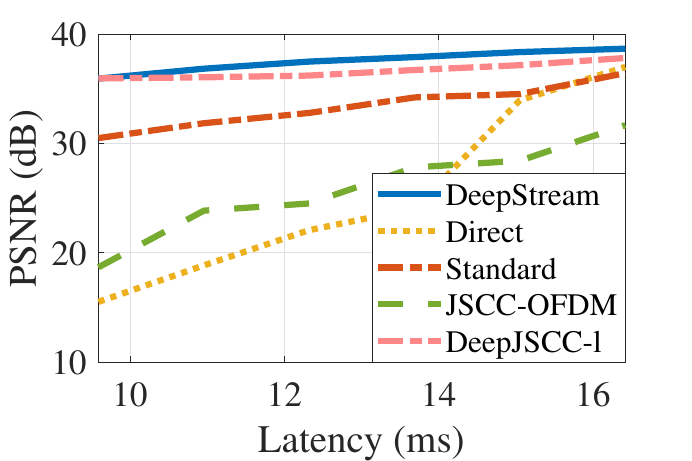}
        \vspace{-1.5em}
        \caption{SNR=20~\!dB.}
        \label{fig:latency_high_snr}
        \includegraphics[width=1\linewidth]{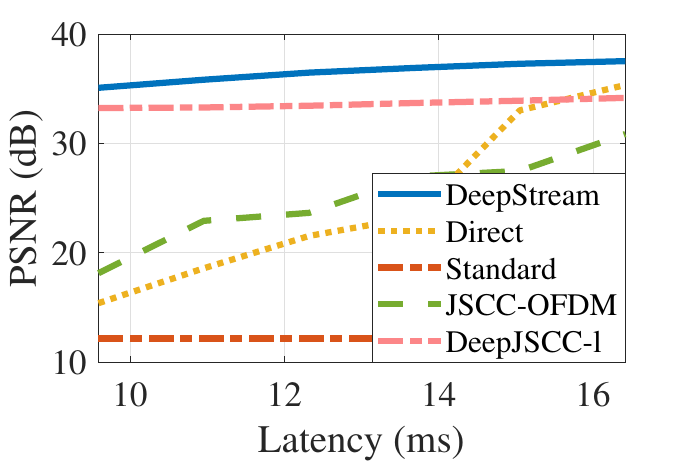}
        \vspace{-1.5em}
        \caption{SNR=12~\!dB.}
        \label{fig:latency_low_snr}
    \end{subfigure}
    \caption{Performance of (a) image transmission with Caltech 101 dataset and (b) video transmission with UCF-101 dataset, (c) and (d) an image example under \name and direct deployment scheme, and performance for image transmission under different latency constraints at (e) high and (f) low SNRs.}
    \label{fig:exp_2}
    \vspace{-1.8em}
\end{figure*}

\subsubsection{Generalization performance}

To verify the generalizability of \name, we conduct experiments from both channel and dataset perspectives.
We first evaluate the performance of \name in the NLoS scenario. To assess whether the trained encoder and decoder generalize beyond specific channel conditions, we conduct image transmission and video streaming experiments in the NLoS scenario. 
Fig.~\ref{fig:NLoS_image} and Fig.~\ref{fig:NLoS_video} present the results. In the NLoS scenario, all schemes demonstrate performance degradation compared with the results in the LoS scenario, but \name still outperforms baselines and achieves performance comparable to its results in the LoS scenario, as shown in Fig.~\ref{fig:end2end_image} and Fig.~\ref{fig:end2end_video}. These results demonstrate that the proposed scheme maintains robust performance across diverse channel conditions.
 
We further evaluate the proposed scheme on unseen datasets. For image transmission, we train the model on the ImageNet dataset~\cite{deng2009imagenet} and evaluate on the Kodak dataset~\cite{kodak}.
For video streaming, we train the model on the Vimeo-90K dataset~\cite{xue2019video} and evaluate on the HEVC Class D dataset~\cite{sullivan2012overview}.
Fig.~\ref{fig:general_image} presents the image transmission performance in the NLoS scenario. Across different SNR levels, \name consistently outperforms the competing schemes on the unseen dataset. Fig.~\ref{fig:general_video} presents the video streaming performance, where \name achieves superior performance over the baselines in the low-SNR regime and consistently outperforms the direct deployment scheme across all SNR levels.
This indicates that even when using the pre-trained model and the precomputed precoding matrix, the proposed scheme still performs well on unseen data, validating its generalizability.

Furthermore, to further validate the generalization capability of \name, we conduct additional experiments on another pair of unseen datasets for both image and video transmission. Specifically, for image transmission, we adopt the Caltech-101 dataset~\cite{fei2004learning}, which contains images from 102 categories, and resize all images to 224$\times$224. For video transmission, we utilize the UCF-101 dataset, where frames are randomly cropped to 256$\times$256 pixels. The evaluation is conducted under the NLoS scenario, with the results shown in Fig.~\ref{fig:general_2_image} and Fig.~\ref{fig:general_2_video}. The results show that \name continues to generalize well across both datasets and consistently outperforms the direct deployment scheme under all SNR conditions, further demonstrating the effectiveness and robustness of the proposed framework.

To further investigate performance, Fig.~\ref{fig:example_1} and Fig.~\ref{fig:example_2} present image transmission results of both the direct deployment scheme and \name for the same image at an SNR of about 10~\!dB. In the direct deployment scheme, noticeable distortions appear in the lower regions, where details deviate significantly from the original image. Additionally, significant blurring occurs in the upper right corner. In contrast, the image reconstructed by \name closely resembles the original, achieving a higher PSNR, thereby demonstrating its effectiveness.

\begin{figure*}[t]
    \centering
    \setlength{\abovecaptionskip}{6pt}
    \begin{subfigure}[b]{0.32\linewidth}
        \centering
        \includegraphics[width=1\linewidth]{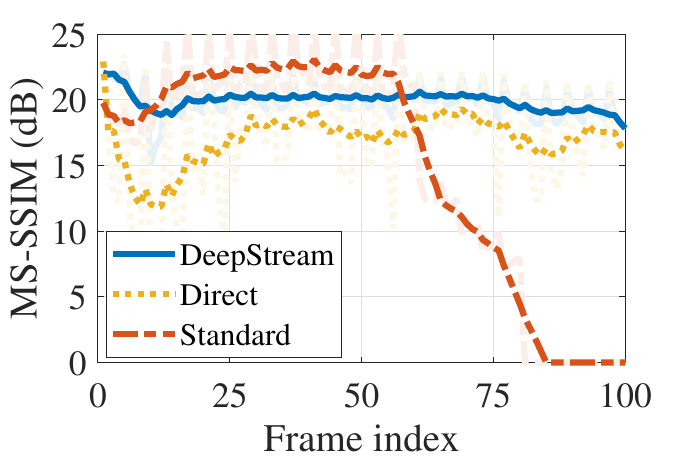}
        \vspace{-1.5em}
        \caption{MS-SSIM.}
        \label{fig:mobile_rt}
        \includegraphics[width=1\linewidth]{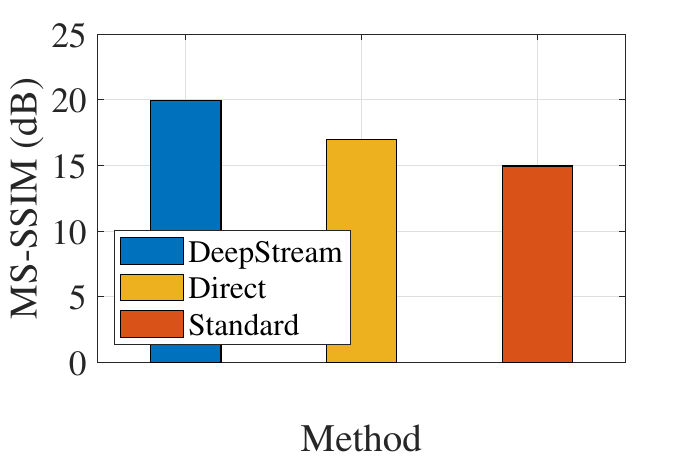}
        \vspace{-1.5em}
        \caption{Average MS-SSIM.}
        \label{fig:mobile_bar}
    \end{subfigure}
    \hfill
    \begin{subfigure}[b]{0.32\linewidth}
        \centering
        \includegraphics[width=1\linewidth]{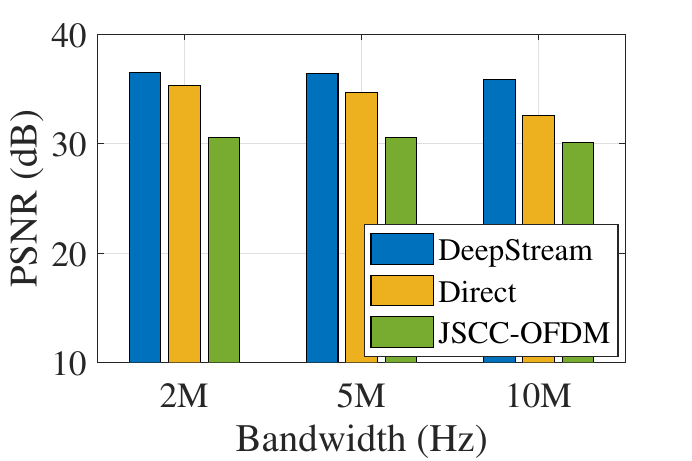}
        \vspace{-1.5em}
        \caption{Various bandwidths.}
        \label{fig:var_bw}
        \includegraphics[width=1\linewidth]{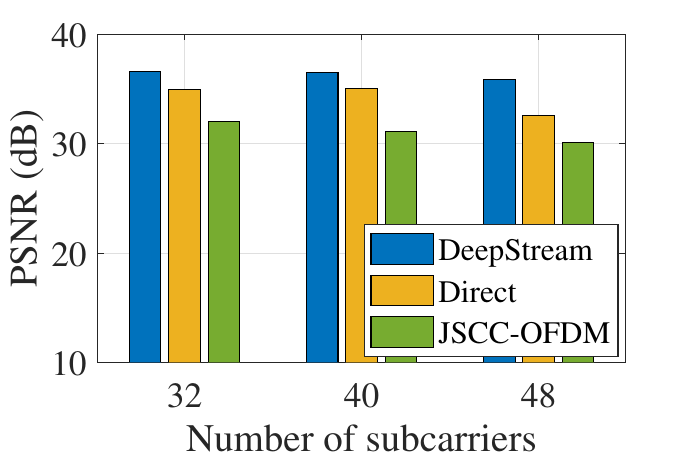}
        \vspace{-1.5em}
        \caption{Various subcarrier counts.}
         \label{fig:var_subcarriers}
    \end{subfigure}
    \hfill
    \begin{subfigure}[b]{0.32\linewidth}
        \centering
        \includegraphics[width=1\linewidth]{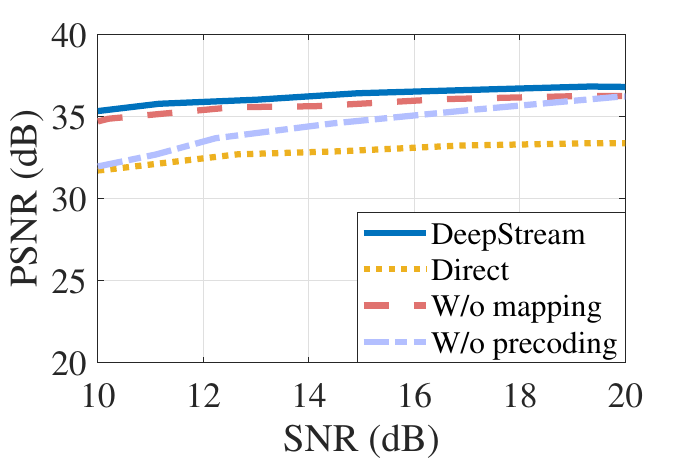}
        \vspace{-1.5em}
        \caption{Performance of each scheme.}
        \label{fig:ablation_image}
        \includegraphics[width=1\linewidth]{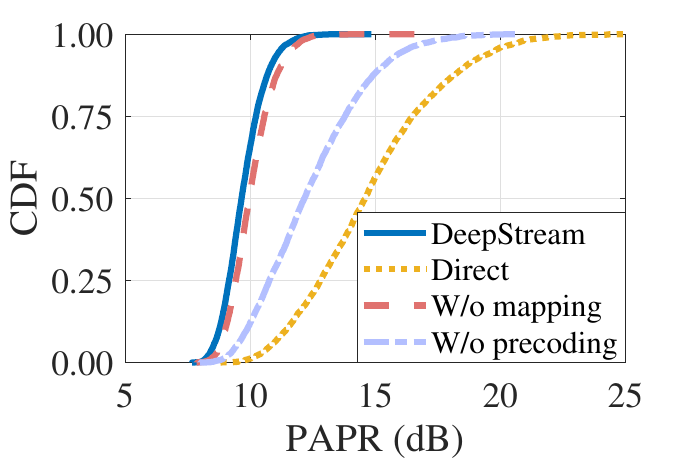}
        \vspace{-1.5em}
        \caption{PAPR (dB).}
        \label{fig:ablation_papr}
    \end{subfigure}
    \caption{Performance for video streaming with mobility: (a) MS-SSIM of each frame and (b) average MS-SSIM, (c) performance under various bandwidths, (d) performance under various subcarrier counts, and ablation study of the proposed components: (e) overall performance and (f) PAPR comparison.}
    \label{fig:exp_2}
    \vspace{-1.8em}
\end{figure*}

\subsubsection{Performance with varying latency}

Fig.~\ref{fig:latency_high_snr} and Fig.~\ref{fig:latency_low_snr} present the image transmission performance under varying latency constraints.
At high SNRs in Fig.~\ref{fig:latency_high_snr}, all schemes experience performance degradation as the latency constraint tightens due to reduced data transmission capacity. Notably, \name shows a significantly slower decline compared to the direct deployment scheme. This is because the progressive encoding strategy effectively prioritizes important information, making it better suitable for urgent transmission requirements. Moreover, \name consistently outperforms DeepJSCC-l (mentioned in Section~\ref{mico-benchmark}) under both settings, highlighting the effectiveness of its fine-grained progressive training strategy. We also observe that JSCC-OFDM experiences significant performance degradation in both scenarios due to the clipping operation. 
Furthermore, \name consistently outperforms the standard scheme, demonstrating the advantages of DeepJSCC in leveraging DL for efficient compression and reconstruction using critical information. 

Meanwhile, at low SNRs in Fig.~\ref{fig:latency_low_snr}, performance degradation is more pronounced due to poor channel conditions. The standard scheme undergoes a substantial performance drop and, even under loose latency constraints, fails to decode correctly, resulting in only minimal PSNR.
In contrast, DeepJSCC-based schemes (\name and direct deployment) sustain relatively high performance, benefiting from their inherent robustness and superior error correction capabilities.

\subsubsection{Performance with mobility}

To further assess the effectiveness of \name in dynamic environments, we evaluate its performance in a mobile scenario. 
Specifically, we first fix the location of the transmitter and predefined two receiver locations corresponding to approximately 20 dB and 12 dB SNR, respectively, and then selected three intermediate locations between them to obtain a set of locations with progressively decreasing SNR levels. At each of the five locations, we transmitted 20 frames of a video sequence. The reported results are obtained by averaging over ten video sequences. The blurred curves represent the averaged measurements, while the solid curves denote the corresponding fitted trends for better visualization.
The corresponding results for video streaming are shown in Fig.~\ref{fig:mobile_rt} and Fig.~\ref{fig:mobile_bar}.
As shown in the results, all three schemes undergo performance degradation as the receiver moves, primarily due to the decreasing SNR. When the SNR is high, the conventional digital scheme achieves better performance because it can effectively mitigate channel noise through channel coding. However, as the receiver moves farther away from the transmitter and the SNR decreases, the conventional scheme becomes highly sensitive to channel fluctuations, resulting in significant performance degradation. 
In contrast, both \name and the direct deployment scheme maintain more stable performance, demonstrating robustness against mobility and fluctuating channel conditions. Furthermore, \name surpasses the direct deployment scheme, further validating the effectiveness of the proposed module.

\subsection{Impact Factors}
To further investigate the performance of \name, we evaluate its robustness under different practical system configurations and analyze the contribution of each proposed module. Specifically, we study the impact of bandwidth and subcarrier count, followed by an ablation study to examine the effectiveness of individual components.

\subsubsection{Impact of Bandwidth} Fig.~\ref{fig:var_bw} presents the results under different bandwidth settings at an SNR of approximately 14 dB in an NLoS scenario. The results show that the performance of all schemes degrades as the bandwidth increases, since a wider bandwidth makes multipath effects more resolvable and thus increases channel distortion. Nevertheless, \name consistently outperforms the baselines and exhibits negligible performance degradation, demonstrating strong generalization across different bandwidth settings.

\subsubsection{Impact of Subcarrier Count} Fig.~\ref{fig:var_subcarriers} presents the performance under different numbers of subcarriers. Specifically, we set the subcarriers far from the direct current carrier to zero in order to control the number of active subcarriers. We observe that the performance of all schemes improves as the number of subcarriers decreases. This is because fewer subcarriers lead to lower PAPR and less pronounced multipath effects, resulting in reduced signal distortion. Meanwhile, \name consistently outperforms the baselines and maintains high performance, demonstrating its robustness across different subcarrier configurations.

\subsubsection{Ablation Study} We further conduct an ablation study to examine the effectiveness of each component under the NLoS scenario, as shown in Fig.~\ref{fig:ablation_image} and Fig.~\ref{fig:ablation_papr}. Specifically, \textit{w/o mapping} denotes the scheme without the proposed feature-to-symbol mapping module, and \textit{w/o precoding} denotes the scheme without the proposed precoding module. For a fair comparison, both neural networks were retrained from scratch. From Fig.~\ref{fig:ablation_image}, we observe that \name consistently outperforms both ablated variants, while both variants still outperform the direct deployment scheme. This demonstrates that both the proposed mapping and precoding modules contribute to the overall performance improvement. Moreover, the scheme without mapping performs better than the scheme without precoding, indicating that the proposed precoding module contributes more significantly to the performance gain. Furthermore, Fig.~\ref{fig:ablation_papr} shows the PAPR of each scheme. We observe that the PAPR of both ablated variants is higher than that of \name, yet still lower than that of the direct deployment scheme. This confirms that both components help reduce PAPR, with the precoding module providing a more pronounced reduction.

\section{Related work and discussion} 
\label{sec:related_work}



\subsection{Related Work}
Recently, DL-based JSCC has been explored as an effective and robust transmission technique~\cite{bourtsoulatze2019deepjscc,yang2022jsccofdm,xu2021wireless,kurka2021bandwidth,wang2022wirelessvideo,tung2022deepwive, zhang2024unified,tung2022deepjscc, chi2026deepguard}. 
Bourtsoulatze~\textit{et al.}~\cite{bourtsoulatze2019deepjscc} proposed DeepJSCC, a CNN-based JSCC scheme optimized for image transmission. The results demonstrate that DeepJSCC achieves robust performance under low SNR conditions, effectively avoiding the cliff effect observed in traditional approaches. 
Xu~\textit{et al.}~\cite{xu2021wireless} extended DeepJSCC using an attention mechanism, incorporating SNR as input to enhance both encoding and decoding processes, making it maintain high performance under diverse SNR conditions.
To enhance compatibility with existing digital communication systems, Tung~\textit{et al.}~\cite{tung2022deepjscc} proposed DeepJSCC-Q, which maps features to the nearest constellation points. 

To enhance the flexibility, some prior works have proposed DeepJSCC schemes with dynamic compression ratios for image transmission~\cite{kurka2021bandwidth, yang2022deep, dai2022nonlinear, zhang2023predictive, bian2023deepjscc, yang2024swinjscc}. However, the approaches in~\cite{yang2022deep, dai2022nonlinear, zhang2023predictive, bian2023deepjscc, yang2024swinjscc} rely on additional side information, such as SNR estimates or entropy-related signals. Delivering such information requires either a CSI feedback link or an auxiliary digital transmission link, which substantially increases system implementation complexity and is beyond the scope of our target system. For example, \cite{yang2022deep, zhang2023predictive, bian2023deepjscc} requires SNR to guide the generation of mask vector or compression rate optimization, and~\cite{dai2022nonlinear} requires extra side information to achieve adaptive transmission.
Although~\cite{kurka2021bandwidth} does not require transmitting additional side information, its adaptation mechanism is relatively coarse-grained. It considers only two layers (each containing multiple channels) with a single decoder and trains the network by randomly masking the refinement layer. This coarse-grained strategy inevitably leads to performance degradation.
Our strategy does not require side information and works in a more fine-granted manner, where the number of masked channels is more flexible.

Meanwhile, several studies have extended DeepJSCC to video transmission and demonstrated superior reconstruction performance under low-SNR conditions~\cite{wang2022wirelessvideo, tung2022deepwive, du2025object, zhang2026bi}. However, these works primarily focus on improving reconstruction quality through sophisticated neural architectures and are mostly evaluated under idealized channel models such as AWGN or flat-fading channels. As a result, practical challenges arising in OFDM systems, including high PAPR, correlated feature transmission, and multipath-induced distortion, are not considered. Furthermore, end-to-end latency optimization for real-time video streaming remains largely unexplored. In contrast, our work focuses on practical OFDM deployment and introduces several system-level optimizations, including a dual-process streaming framework and quantization techniques, to reduce processing delay while improving transmission robustness in real-world wireless environments.

Besides, above studies primarily focus on AWGN channels, which are different from real-world wireless environments.
To address the issue, Yang~\textit{et al.}~\cite{yang2022jsccofdm} and Shao~\textit{et al.}~\cite{shao2022semanticofdm} extended JSCC to multipath channels, while Wu~\textit{et al.}~\cite{wu2024jsccmimo} applied it to multiple-input multiple-output systems. 
Yoo~\cite{yoo2023role} prototyped a ViT-based DeepJSCC with USRP for image transmission, but they dit not consider the OFDM configuration.
Liu~\textit{et al.}~\cite{liu2022real} proposed an SDR-based image transmission framework based on OFDM, but did not consider the effect of the multipath fading channel, which is the same as the direct deployment scheme described in experiment. 
In parallel, growing research efforts have explored resource allocation and system management strategies tailored for DeepJSCC systems to further enhance their deployment efficiency~\cite{mu2022heterogeneous, liu2024ofdm, wang2022performance}. 


\subsection{Discussion}
While most existing works show promising results, their evaluations remain limited to simulations, leaving real-world feasibility uncertain. To bridge this gap, we introduce \name, a prototype system for DeepJSCC.
Our primary objective is to integrate DeepJSCC with OFDM systems and fully exploit its potential. 
We focus on intermediate-layer designs that effectively align the JSCC encoding process with OFDM transmission. We adopt a modified activation function that restricts the range of the output of the encoder to reduce PAPR. While this design demonstrates clear empirical benefits in practical OFDM transmission, a rigorous theoretical characterization of its impact on information loss and rate-distortion performance remains an interesting direction for future work. Likewise, the proposed precoding algorithm relies on a computationally efficient suboptimal solution. Although it performs well in practice, establishing theoretical optimality guarantees remains an important topic for future research.
Beside, although we do not modify the encoder or decoder directly, our system remains compatible with a wide range of existing neural network architectures.
For instance, recent advances such as vision transformers (ViT)~\cite{wu2024jsccmimo} and diffusion models~\cite{guo2025diffusion}, have been employed to further improve the transmission efficiency of DeepJSCC systems in image transmission. 
Similarly, although \name may exhibit inferior performance compared with advanced video codecs in video transmission in high-SNR scenarios, its performance could potentially be further improved by incorporating more advanced neural network architectures. For example, ViT-based architectures have also shown promising performance in video transmission~\cite{wang2022wirelessvideo}.
By integrating the proposed feature-to-symbol mapping method and cross-subcarrier precoding technique, these architectures can be made more compatible with OFDM-based transmission. 
Nevertheless, many of these approaches primarily focus on improving reconstruction performance in simulation environments and often incur considerable computational overhead, which may limit their applicability to real-time streaming scenarios. In addition, some methods rely on extra side information, such as CSI, SNR estimates, or entropy-related information, which requires additional feedback or digital transmission links and further increases system complexity. Therefore, while these architectures may provide additional performance gains, enabling low-latency and practical deployment remains a significant challenge. Future research should focus on developing efficient neural architectures and system designs that achieve a better balance among reconstruction quality, computational complexity, system complexity, and transmission latency for practical wireless multimedia streaming.
Lastly, as a pioneering effort, our work primarily addresses latency and technical challenges at the physical layer. Nevertheless, our framework can be seamlessly extended to broader deployment scenarios by incorporating network-induced latency considerations~\cite{fouladi2018salsify} and additional technologies (e.g., multiple-input and multiple-output~\cite{wu2024jsccmimo} and cooperative relay networks~\cite{bian2025process}). Besides, this work does not include a dedicated design specifically targeting high-mobility scenarios. Nevertheless, our experimental results demonstrate that the proposed system operates reliably when the devices are not fixed, which can be largely attributed to the inherent robustness of OFDM system. For scenarios involving higher mobility that exceed the capability of OFDM, prior work has explored advanced waveforms such as orthogonal time-frequency space (OTFS) for semantic communication~\cite{chen2024semantic}. Investigating such extensions is beyond the scope of this paper and is left for future work.

\section{Conclusion} \label{sec:conclusion}
In this paper, we have prototyped \name, a real-time noise-resilient DeepJSCC system designed for efficient and robust multimedia transmission. To enhance the compatibility of DeepJSCC with OFDM systems, we introduce a feature-to-symbol mapping method and a cross-subcarrier precoding technique. Additionally, we develop a progressive coding strategy to improve transmission flexibility under varying latency constraints.
Extensive evaluations have demonstrated that \name consistently outperforms baseline methods across a wide range of SNRs, particularly in low-SNR conditions. By bridging the gap between theoretical advancements and practical implementation, \name takes a significant step toward the real-world deployment of DeepJSCC systems.

\bibliographystyle{IEEEtran}
\bibliography{main}

@article{duan20236g,
  title={6G Architecture Design: from Overall, Logical and Networking Perspective},
  author={Duan, Xiao Dong and Wang, Xiao Yun and Lu, Lu and Shi, Nan Xiang and Liu, Chao and Zhang, Tong and Sun, Tao},
  journal={IEEE Commun. Mag.},
  volume={61},
  number={7},
  pages={158--164},
  year={2023},
  publisher={IEEE}
}

@article{chen20235g,
  title={5G-advanced toward 6G: Past, present, and future},
  author={Chen, Wanshi and Lin, Xingqin and Lee, Juho and Toskala, Antti and Sun, Shu and Chiasserini, Carla Fabiana and Liu, Lingjia},
  journal={IEEE J. Sel. Areas Commun.},
  volume={41},
  number={6},
  pages={1592--1619},
  year={2023},
  publisher={IEEE}
}

@article{wang2023road,
  title={On the road to 6G: Visions, requirements, key technologies, and testbeds},
  author={Wang, Cheng-Xiang and You, Xiaohu and Gao, Xiqi and Zhu, Xiuming and Li, Zixin and Zhang, Chuan and Wang, Haiming and Huang, Yongming and Chen, Yunfei and Haas, Harald and others},
  journal={IEEE Commun. Surveys Tuts.},
  volume={25},
  number={2},
  pages={905--974},
  year={2023},
  publisher={IEEE}
}

@article{liu2024road,
  title={The road to next-generation multiple access: A 50-year tutorial review},
  author={Liu, Yuanwei and Ouyang, Chongjun and Ding, Zhiguo and Schober, Robert},
  journal={Proc. IEEE},
  year={2024},
  publisher={IEEE}
}

@article{erdemir2023generative,
  title={Generative Joint Source-Channel Coding for Semantic Image Transmission},
  author={Erdemir, Ecenaz and Tung, Tze-Yang and Dragotti, Pier Luigi and G{\"u}nd{\"u}z, Deniz},
  journal={IEEE J. Sel. Areas Commun.},
  volume={41},
  number={8},
  pages={2645--2657},
  year={2023},
  publisher={IEEE}
}

@misc{3gpp22125,
  author = {{3GPP}},
  title = {{{Unmanned Aerial System (UAS) support in 3GPP}}},
  howpublished = {\url{http://www.3gpp.org/DynaReport/22125.htm}},
  note = {Online; accessed: 16 January 2025},
  year = 2024
}

@inproceedings{ballé2018variational,
  title={Variational Image Compression with a Scale Hyperprior},
  author={Johannes Ballé and David Minnen and Saurabh Singh and Sung Jin Hwang and Nick Johnston},
  booktitle={Proc. Int. Conf. Learn. Represent. (ICLR)},
  pages = {1--23},
  year={2018},
}

@article{shao2021learning,
  title={Learning task-oriented communication for edge inference: An information bottleneck approach},
  author={Shao, Jiawei and Mao, Yuyi and Zhang, Jun},
  journal={IEEE J. Sel. Areas Commun.},
  volume={40},
  number={1},
  pages={197--211},
  year={2021},
  publisher={IEEE}
}

@inproceedings{li2021deepvideo,
  title={Deep Contextual Video Compression},
  author={Li, Jiahao and Li, Bin and Lu, Yan},
  booktitle={Adv. Neural Inf. Process. Syst. (NeurIPS)},
  pages = {18114--18125},
  year={2021}
}

@inproceedings{du2020server,
  title={Server-Driven Video Streaming for Deep Learning Inference},
  author={Du, Kuntai and Pervaiz, Ahsan and Yuan, Xin and Chowdhery, Aakanksha and Zhang, Qizheng and Hoffmann, Henry and Jiang, Junchen},
  booktitle={Proc. ACM SIGCOMM Conf.},
  pages={557--570},
  year={2020}
}

@article{gunduz2022beyond,
  title={Beyond Transmitting Bits: Context, Semantics, and Task-Oriented Communications},
  author={G{\"u}nd{\"u}z, Deniz and Qin, Zhijin and Aguerri, Inaki Estella and Dhillon, Harpreet S and Yang, Zhaohui and Yener, Aylin and Wong, Kai Kit and Chae, Chan-Byoung},
  journal={IEEE J. Sel. Areas Commun.},
  volume={41},
  number={1},
  pages={5--41},
  year={2022},
  publisher={IEEE}
}

@article{luo2022semantic,
  title={Semantic Communications: Overview, Open Issues, and Future Research Directions},
  author={Luo, Xuewen and Chen, Hsiao-Hwa and Guo, Qing},
  journal={IEEE Wireless Comm.},
  volume={29},
  number={1},
  pages={210--219},
  year={2022},
  publisher={IEEE}
}

@article{bourtsoulatze2019deepjscc,
  title={Deep Joint Source-Channel Coding for Wireless Image Transmission},
  author={Bourtsoulatze, Eirina and Kurka, David Burth and G{\"u}nd{\"u}z, Deniz},
  journal={IEEE Trans. Cogn. Commun. Netw.},
  volume={5},
  number={3},
  pages={567--579},
  year={2019},
  publisher={IEEE}
}

@article{xu2021wireless,
  title={Wireless Image Transmission Using Deep Source Channel Coding With Attention Modules},
  author={Xu, Jialong and Ai, Bo and Chen, Wei and Yang, Ang and Sun, Peng and Rodrigues, Miguel},
  journal={IEEE Trans. Circuits Syst. Video Technol.},
  volume={32},
  number={4},
  pages={2315--2328},
  year={2021},
  publisher={IEEE}
}

@article{huang2022toward,
  title={Toward Semantic Communications: Deep Learning-Based Image Semantic Coding},
  author={Huang, Danlan and Gao, Feifei and Tao, Xiaoming and Du, Qiyuan and Lu, Jianhua},
  journal={IEEE J. Sel. Areas Commun.},
  volume={41},
  number={1},
  pages={55--71},
  year={2022},
  publisher={IEEE}
}

@article{han2022semanticspeech,
  title={Semantic-preserved Communication System for Highly Efficient Speech Transmission},
  author={Han, Tianxiao and Yang, Qianqian and Shi, Zhiguo and He, Shibo and Zhang, Zhaoyang},
  journal={IEEE J. Sel. Areas Commun.},
  volume={41},
  number={1},
  pages={245--259},
  year={2022},
  publisher={IEEE}
}

@article{weng2021semanticspeech,
  title={Semantic Communication Systems for Speech Transmission},
  author={Weng, Zhenzi and Qin, Zhijin},
  journal={IEEE J. Sel. Areas Commun.},
  volume={39},
  number={8},
  pages={2434--2444},
  year={2021},
  publisher={IEEE}
}

@article{tung2022deepwive,
  title={DeepWiVe: Deep-Learning-Aided Wireless Video Transmission},
  author={Tung, Tze-Yang and G{\"u}nd{\"u}z, Deniz},
  journal={IEEE J. Sel. Areas Commun.},
  volume={40},
  number={9},
  pages={2570--2583},
  year={2022},
  publisher={IEEE}
}

@article{wang2022wirelessvideo,
  title={Wireless Deep Video Semantic Transmission},
  author={Wang, Sixian and Dai, Jincheng and Liang, Zijian and Niu, Kai and Si, Zhongwei and Dong, Chao and Qin, Xiaoqi and Zhang, Ping},
  journal={IEEE J. Sel. Areas Commun.},
  volume={41},
  number={1},
  pages={214--229},
  year={2022},
  publisher={IEEE}
}

@article{yang2022jsccofdm,
  title={OFDM-guided Deep Joint Source Channel Coding for Wireless Multipath Fading Channels},
  author={Yang, Mingyu and Bian, Chenghong and Kim, Hun-Seok},
  journal={IEEE Trans. Cogn. Commun. Netw.},
  volume={8},
  number={2},
  pages={584--599},
  year={2022},
  publisher={IEEE}
}

@article{shao2022semanticofdm,
  title={Semantic Communications with Discrete-time Analog Transmission: A PAPR Perspective},
  author={Shao, Yulin and Gunduz, Deniz},
  journal={IEEE Wireless Commun. Lett.},
  volume={12},
  number={3},
  pages={510--514},
  year={2022},
  publisher={IEEE}
}

@article{wu2024jsccmimo,
  title={Deep Joint Source-Channel Coding for Adaptive Image Transmission over MIMO Channels},
  author={Wu, Haotian and Shao, Yulin and Bian, Chenghong and Mikolajczyk, Krystian and G{\"u}nd{\"u}z, Deniz},
  journal={IEEE Trans. Wireless Commun.},
  volume={23},
  number={10},
  pages={15002--15017},
  year={2024},
  publisher={IEEE}
}

@article{mu2022heterogeneous,
  title={Heterogeneous Semantic and Bit Communications: A Semi-NOMA Scheme},
  author={Mu, Xidong and Liu, Yuanwei and Guo, Li and Al-Dhahir, Naofal},
  journal={IEEE J. Sel. Areas Commun.},
  volume={41},
  number={1},
  pages={155--169},
  year={2022},
  publisher={IEEE}
}

@article{chi2024capacity,
  title={Capacity Optimizing Resource Allocation in Joint Source-Channel Coding Systems with QoS Constraints},
  author={Chi, Kaiyi and Yang, Qianqian and Yang, Zhaohui and Duan, Yiping and Zhang, Zhaoyang},
  journal={IEEE Trans. Commun.},
  volume={73},
  number={6},
  pages={4198-4212},
  year={2025},
  publisher={IEEE}
}

@article{bolcskei2003impact,
  title={Impact of the Propagation Environment on the Performance of Space-Frequency Coded MIMO-OFDM},
  author={Bolcskei, H and Borgmann, Moritz and Paulraj, Arogyaswami J},
  journal={IEEE J. Sel. Areas Commun.},
  volume={21},
  number={3},
  pages={427--439},
  year={2003},
  publisher={IEEE}
}

@inproceedings{park2000papr,
  title={PAPR Reduction in OFDM Transmission using Hadamard Transform},
  author={Park, Myonghee and Jun, Heeyoung and Cho, Jaehee and Cho, Namshin and Hong, Daesik and Kang, Changeun},
  booktitle={Proc. IEEE Int. Conf. Commun. (ICC)},
  pages={430--433},
  year={2000}
}

@article{yoo2023role,
  title={On the role of ViT and CNN in semantic communications: Analysis and prototype validation},
  author={Yoo, Hanju and Dai, Linglong and Kim, Songkuk and Chae, Chan-Byoung},
  journal={IEEE Access},
  volume={11},
  pages={71528--71541},
  year={2023},
  publisher={IEEE}
}

@inproceedings{yang2022deep,
  title={Deep joint source-channel coding for wireless image transmission with adaptive rate control},
  author={Yang, Mingyu and Kim, Hun-Seok},
  booktitle={Proc. IEEE Int. Conf. Acoust., Speech, Signal Process. (ICASSP)},
  pages={5193--5197},
  year={2022}
}

@article{kurka2020deepjscc_f,
  title={DeepJSCC-f: Deep Joint Source-Channel Coding of Images with Feedback},
  author={Kurka, David Burth and G{\"u}nd{\"u}z, Deniz},
  journal={IEEE J. Sel. Areas Inf. Theory},
  volume={1},
  number={1},
  pages={178--193},
  year={2020},
  publisher={IEEE}
}

@misc{kodak,
	author = {{Rich Franzen}},
	title = {{{Kodak Lossless True Color Image Suite}}},
	howpublished = {\url{https://r0k.us/graphics/kodak/}},
	note = {Online; accessed: 10 February 2025},
	year = 2025
}

@article{soomro2012ucf101,
  title={UCF101: A dataset of 101 human actions classes from videos in the wild},
  author={Soomro, Khurram and Zamir, Amir Roshan and Shah, Mubarak},
  journal={arXiv preprint arXiv:1212.0402},
  year={2012}
}

@article{xue2019video,
  title={Video Enhancement with Task-Oriented Flow},
  author={Xue, Tianfan and Chen, Baian and Wu, Jiajun and Wei, Donglai and Freeman, William T},
  journal={Int. J. Comput. Vis.},
  volume={127},
  pages={1106--1125},
  year={2019},
  publisher={Springer}
}

@article{sullivan2012overview,
  title={Overview of the High Efficiency Video Coding (HEVC) Standard},
  author={Sullivan, Gary J and Ohm, Jens-Rainer and Han, Woo-Jin and Wiegand, Thomas},
  journal={IEEE Trans. Circuits Syst. Video Technol.},
  volume={22},
  number={12},
  pages={1649--1668},
  year={2012},
  publisher={IEEE}
}

@article{gallager1962low,
  title={Low-Density Parity-Check Codes},
  author={Gallager, Robert},
  journal={IEEE Trans. Inf. Theory},
  volume={8},
  number={1},
  pages={21--28},
  year={1962},
  publisher={IEEE}
}

@article{mackay2005fountain,
  title={Fountain Codes},
  author={MacKay, David JC},
  journal={IEE Proceedings-Communications},
  volume={152},
  number={6},
  pages={1062--1068},
  year={2005},
  publisher={IET}
}

@article{tang2024contrastive,
  title={Contrastive Learning based Semantic Communications},
  author={Tang, Shunpu and Yang, Qianqian and Fan, Lisheng and Lei, Xianfu and Nallanathan, Arumugam and Karagiannidis, George K},
  journal={IEEE Trans. Commun.},
  volume={72},
  number={10},
  pages={6328-6343},
  year={2024},
  publisher={IEEE}
}

@article{kurka2021bandwidth,
  title={Bandwidth-Agile Image Transmission with Deep Joint Source-Channel Coding},
  author={Kurka, David Burth and G{\"u}nd{\"u}z, Deniz},
  journal={IEEE Trans. Wireless Commun.},
  volume={20},
  number={12},
  pages={8081--8095},
  year={2021},
  publisher={IEEE}
}

@article{tung2022deepjscc,
  title={DeepJSCC-Q: Constellation Constrained Deep Joint Source-Channel Coding},
  author={Tung, Tze-Yang and Kurka, David Burth and Jankowski, Mikolaj and G{\"u}nd{\"u}z, Deniz},
  journal={IEEE J. Sel. Areas Inf. Theory},
  volume={3},
  number={4},
  pages={720--731},
  year={2022},
  publisher={IEEE}
}

@article{zhang2024unified,
  title={A Unified Multi-Task Semantic Communication System for Multimodal Data},
  author={Zhang, Guangyi and Hu, Qiyu and Qin, Zhijin and Cai, Yunlong and Yu, Guanding and Tao, Xiaoming},
  journal={IEEE Trans. Commun.},
  volume={72},
  number={7},
  pages={4101--4116},
  year={2024},
  publisher={IEEE}
}

@article{zhang2023predictive,
  title={Predictive and Adaptive Deep Coding for Wireless Image Transmission in Semantic Communication},
  author={Zhang, Wenyu and Zhang, Haijun and Ma, Hui and Shao, Hua and Wang, Ning and Leung, Victor CM},
  journal={IEEE Trans. Wireless Commun.},
  volume={22},
  number={8},
  pages={5486--5501},
  year={2023},
  publisher={IEEE}
}

@article{gunduz2024joint,
  title={Joint Source-Channel coding: Fundamentals and Recent Progress in Practical Designs},
  author={G{\"u}nd{\"u}z, Deniz and Wigger, Mich{\`e}le A and Tung, Tze-Yang and Zhang, Ping and Xiao, Yong},
  journal={Proc. IEEE},
  year={2024},
  publisher={IEEE}
}

@article{dai2022nonlinear,
  title={Nonlinear Transform Source-Channel Coding for Semantic Communications},
  author={Dai, Jincheng and Wang, Sixian and Tan, Kailin and Si, Zhongwei and Qin, Xiaoqi and Niu, Kai and Zhang, Ping},
  journal={IEEE J. Sel. Areas Commun.},
  volume={40},
  number={8},
  pages={2300--2316},
  year={2022},
  publisher={IEEE}
}

@inproceedings{fouladi2018salsify,
  title={Salsify: Low-Latency Network Video through Tighter Integration between a Video Codec and a Transport Protocol},
  author={Fouladi, Sadjad and Emmons, John and Orbay, Emre and Wu, Catherine and Wahby, Riad S and Winstein, Keith},
  booktitle={Proc. USENIX NSDI},
  pages={267--282},
  year={2018}
}

@article{guo2025diffusion,
  title={Diffusion-driven semantic communication for generative models with bandwidth constraints},
  author={Guo, Lei and Chen, Wei and Sun, Yuxuan and Ai, Bo and Pappas, Nikolaos and Quek, Tony},
  journal={IEEE Trans. Wireless Commun.},
  year={2025},
  publisher={IEEE}
}

@article{wang2022performance,
  title={Performance optimization for semantic communications: An attention-based reinforcement learning approach},
  author={Wang, Yining and Chen, Mingzhe and Luo, Tao and Saad, Walid and Niyato, Dusit and Poor, H Vincent and Cui, Shuguang},
  journal={IEEE J. Sel. Areas Commun.},
  volume={40},
  number={9},
  pages={2598--2613},
  year={2022},
  publisher={IEEE}
}

@article{liu2024ofdm,
  title={OFDM-based digital semantic communication with importance awareness},
  author={Liu, Chuanhong and Guo, Caili and Yang, Yang and Ni, Wanli and Quek, Tony QS},
  journal={IEEE Trans. Commun.},
  year={2024},
  publisher={IEEE}
}

@article{bian2025process,
  title={Process-and-forward: Deep joint source-channel coding over cooperative relay networks},
  author={Bian, Chenghong and Shao, Yulin and Wu, Haotian and Ozfatura, Emre and G{\"u}nd{\"u}z, Deniz},
  journal={IEEE J. Sel. Areas Commun.},
  year={2025},
  publisher={IEEE}
}

@inproceedings{he2016deep,
  title={Deep Residual Learning for Image Recognition},
  author={He, Kaiming and Zhang, Xiangyu and Ren, Shaoqing and Sun, Jian},
  booktitle={Proc. IEEE/CVF Conf. Comput. Vis. Pattern Recognit. (CVPR)},
  pages={770--778},
  year={2016}
}

@inproceedings{liu2022real,
  title={Real-time Implementation and Evaluation of SDR-based Deep Joint Source-Channel Coding},
  author={Liu, Maolin and Chen, Wei and Xu, Jialong and Ai, Bo},
  booktitle={Proc. IEEE Veh. Technol. Conf. (VTC)},
  pages={1--5},
  year={2022}
}

@inproceedings{koike2020stochastic,
  title={Stochastic Bottleneck: Rateless Auto-Encoder for Flexible Dimensionality Reduction},
  author={Koike-Akino, Toshiaki and Wang, Ye},
  booktitle={2020 IEEE International Symposium on Information Theory (ISIT)},
  pages={2735--2740},
  year={2020}
}

@inproceedings{hanin2018neural,
  title={Which Neural Net Architectures Give Rise to Exploding and Vanishing Gradients?},
  author={Hanin, Boris},
  booktitle={Adv. Neural Inf. Process. Syst. (NeurIPS)},
  pages={1--10},
  year={2018}
}

@book{terry2002ofdm,
  title={OFDM Wireless LANs: A Theoretical and Practical Guide},
  author={Terry, John and Heiskala, Juha},
  year={2002},
  publisher={Sams publishing}
}

@misc{usrpx310,
	author = {{Ettus Research}},
	title = {{{USRP X300/X310}}},
	howpublished = {\url{https://kb.ettus.com/X300/X310}},
	note = {Online; accessed: 16 February 2025},
	year = 2025
}

@misc{uhd,
	author = {{Ettus Research}},
	title = {{{USRP Hardware Driver and USRP Manual}}},
	howpublished = {\url{https://files.ettus.com/manual/}},
	note = {Online; accessed: 16 February 2025},
	year = 2025
}

@inproceedings{deng2009imagenet,
  title={ImageNet: A Large-Scale Hierarchical Image Database},
  author={Deng, Jia and Dong, Wei and Socher, Richard and Li, Li-Jia and Li, Kai and Fei-Fei, Li},
  booktitle={Proc. IEEE/CVF Conf. Comput. Vis. Pattern Recognit. (CVPR)},
  pages={248--255},
  year={2009}
}

@book{tse2005fundamentals,
  title={Fundamentals of Wireless Communication},
  author={Tse, David and Viswanath, Pramod},
  year={2005},
  publisher={Cambridge university press}
}

@misc{CVX,
	author = {{CVX Research}},
	title = {{CVX: Matlab Software for Disciplined Convex Programming}},
	howpublished = {\url{https://cvxr.com/cvx/}},
	note = {Online; accessed: 16 February 2025},
	year = 2025
}

@article{bezdek2003convergence,
  title={Convergence of Alternating Optimization},
  author={Bezdek, James C and Hathaway, Richard J},
  journal={Neural, Parallel \& Scientific Computations},
  volume={11},
  number={4},
  pages={351--368},
  year={2003},
  publisher={Dynamic Publishers, Inc. Atlanta, GA, USA}
}

@techreport{3GPP_38.901_R14,
  author = {{3rd Generation Partnership Project (3GPP)}},
  title = {Study on Channel Model for Frequencies from 0.5 to 100 GHz (Release 14)},
  institution = {3GPP},
  number = {3GPP TR 38.901},
  year = {2017},
  url = {https://www.3gpp.org/ftp/Specs/archive/38_series/38.901/38901-100.zip},
  note = {Online; accessed: 16 February 2025}
}

@book{li2006orthogonal,
  title={Orthogonal Frequency Division Multiplexing for Wireless Communications},
  author={Li, Ye Geoffrey and Stuber, Gordon L},
  year={2006},
  publisher={Springer Science \& Business Media}
}

@article{crow1997ieee,
  title={IEEE 802.11 Wireless Local Area Networks},
  author={Crow, Brian P and Widjaja, Indra and Kim, Jeong Geun and Sakai, Prescott T},
  journal={IEEE Commun. Mag.},
  volume={35},
  number={9},
  pages={116--126},
  year={1997},
  publisher={IEEE}
}

@inproceedings{jain20sec, 
  author={Samvit Jain and Xun Zhang and Yuhao Zhou and Ganesh Ananthanarayanan and Junchen Jiang and Yuanchao Shu and Paramvir Bahl and Joseph Gonzalez}, 
  booktitle={Proc. ACM/IEEE Symp. Edge Comput. (SEC)}, 
  title={{Spatula: Efficient Cross-camera Video Analytics on Large Camera Networks}}, 
   pages={110--124},
  year={2020},
}

@inproceedings{jiang21mobicom, 
  author={Jiang, Shiqi and Lin, Zhiqi and Li, Yuanchun and Shu, Yuanchao and Liu, Yunxin}, 
  booktitle={Proc. ACM Int. Conf. Mobile Comput. Netw. (MobiCom)}, 
  title={{Flexible High-resolution Object Detection on Edge Devices with Tunable Latency}}, 
  pages = {559–572},
  year={2021},
}

@inproceedings{lu23aaai, 
  author={Yan Lu and Zhun Zhong and Yuanchao Shu}, 
  booktitle={Proc. AAAI Conf. Artif. Intell. (AAAI)}, 
  title={{Multi-View Domain Adaptive Object Detection in Surveillance Cameras}}, 
  pages={8966--8974},
  year={2023},
}

@inproceedings{li21sec, 
  author={Zhuqi Li and Yuanchao Shu and Ganesh Ananthanarayanan and Longfei Shangguan and Kyle Jamieson and Paramvir Bahl}, 
  booktitle={Proc. ACM/IEEE Symp. Edge Comput. (SEC)}, 
  title={{Spider: A Multi-Hop Millimeter-Wave Network for Live Video Analytics}}, 
  pages={178--191},
  year={2021},
}

@article{chi2026deepguard,
  title={DeepGuard: Defending Deep Joint Source-Channel Coding Against Eavesdropping at Physical-Layer},
  author={Chi, Kaiyi and He, Yinghui and Yang, Qianqian and Shu, Yuanchao and Wang, Zhiqin and Luo, Jun and Chen, Jiming},
  journal={IEEE J. Sel. Areas Commun.},
  year={2026},
  volume={44},
  number={},
  pages={4128--4143},
  publisher={IEEE}
}

@article{yang2024swinjscc,
  title={Swinjscc: Taming swin transformer for deep joint source-channel coding},
  author={Yang, Ke and Wang, Sixian and Dai, Jincheng and Qin, Xiaoqi and Niu, Kai and Zhang, Ping},
  journal={IEEE Trans. Cogn. Commun. Netw.},
  volume={11},
  number={1},
  pages={90--104},
  year={2024},
  publisher={IEEE}
}

@inproceedings{bian2023deepjscc,
  title={DeepJSCC-1++: Robust and bandwidth-adaptive wireless image transmission},
  author={Bian, Chenghong and Shao, Yulin and G{\"u}nd{\"u}z, Deniz},
  booktitle={Proc. IEEE Glob. Commun. Conf. (GLOBECOM)},
  pages={3148--3154},
  year={2023}
}

@article{du2025object,
  title={Object-attribute-relation representation based video semantic communication},
  author={Du, Qiyuan and Duan, Yiping and Yang, Qianqian and Tao, Xiaoming and Debbah, M{\'e}rouane},
  journal={IEEE J. Sel. Areas Commun.},
  year={2025},
  volume={43},
  number={7},
  pages={2446-2461},
  publisher={IEEE}
}

@article{zhang2026bi,
  title={Bi-Directional Motion Enhanced Semantic Communication for Wireless Video Transmission},
  author={Zhang, Zhenguo and Yang, Qianqian and He, Shibo and Shi, Zhiguo},
  journal={IEEE Internet Things J.},
  year={2026},
  volume={13},
  number={8},
  pages={15607-15620},
  publisher={IEEE}
}

@inproceedings{fei2004learning,
  title={Learning generative visual models from few training examples: An incremental bayesian approach tested on 101 object categories},
  author={Fei-Fei, Li and Fergus, Rob and Perona, Pietro},
  booktitle={Proc. IEEE/CVF Conf. Comput. Vis. Pattern Recognit. Workshops (CVPRW)},
  pages={178--178},
  year={2004}
}

@article{chen2024semantic,
  author  = {Chen, Wenbin and Li, Songyan and Ju, Cheng and Wang, Dongdong and Li, Jin and Wang, Danshi},
  title   = {Semantic-enhanced downlink {LEO} satellite communication system with {OTFS} modulation},
  journal = {IEEE Commun. Lett.},
  volume  = {28},
  number  = {6},
  pages   = {1377--1381},
  year    = {2024}
}

@inproceedings{xu2011muzi,
  title={MuZi: Multi-channel ZigBee networks for avoiding WiFi interference},
  author={Xu, Ruitao and Shi, Gaotao and Luo, Jun and Zhao, Zenghua and Shu, Yantai},
  booktitle={Proc. IEEE Int. Conf. Internet Things (iThings) / IEEE Int. Conf. Cyber-Phys. Social Comput. (CPSCom)},
  pages={323--329},
  year={2011},
  organization={IEEE}
}

\end{document}